\documentclass[]{aastex631} 
\usepackage{amssymb}
\usepackage{epsfig}
\usepackage{rotating}
\usepackage{mathtools}
\usepackage{comment}
\usepackage{bm}
\usepackage{epstopdf}
\usepackage{amssymb}
\usepackage{rotating}
\usepackage{array}
\newcolumntype{H}{>{\setbox0=\hbox\bgroup}c<{\egroup}@{}}
\usepackage[caption=false]{subfig}
\newcommand{\kms}{\mbox{km~s$^{-1}$}}
\newcommand{\s}{\mbox{$''$}}

\newcommand{\mdot}{\mbox{$\dot{M}$}}
\newcommand{\my}{\mbox{$M_{\odot}$~yr$^{-1}$}}
\newcommand{\ls}{\mbox{$L_{\odot}$}}
\newcommand{\Zs}{\mbox{Z$_{\odot}$}}

\newcommand{\ms}{\mbox{$M_{\odot}$}}
\newcommand{\me}{M${_\oplus}$}

\newcommand{\bfluxu}{\mbox{erg\,s$^{-1}$\,cm$^{-2}$}}

\newcommand{\densunit}{g\,cm$^{-3}$}

\newcommand{\fbolunit}{\mbox{erg\,s$^{-1}$\,cm$^{-2}$}}

\newcommand{\gsim}{\raisebox{-.4ex}{$\stackrel{>}{\scriptstyle\sim}$}}

\DeclareCaptionFormat{cont}{#1 (cont.)#2#3\par}

\shorttitle{The binary and the disk: the beauty is found within NGC3132 with JWST}
\shortauthors{Sahai et al.}

\begin{document}
\title{The binary and the disk: the beauty is found within NGC3132 with JWST}
\author{Raghvendra Sahai}
\affiliation{Jet Propulsion Laboratory, California Institute of Technology, Pasadena, CA 91109, USA}
\author{Valentin Bujarrabal}
\affiliation{Observatorio Astro\'omico Nacional (OAN/IGN), Ap 112, 28803 Alcal\'a de Henares, Spain}
\author{Guillermo Quintana-Lacaci}
\affiliation{Dept. of Molecular Astrophysics, IFF-CSIC. C/ Serrano 123, E-28006, Madrid, Spain}
\author{Nicole Reindl}
\affiliation{Institut f{\"u}r Physik und Astronomie, Universit{\"a}t Potsdam, Karl-Liebknecht-Stra$\beta$e 24/25, D-14476, Potsdam, Germany}
\author{Griet Van de Steene}
\affiliation{Royal Observatory of Belgium, Astronomy and Astrophysics, Ringlaan 3, 1180 Brussels, Belgium}
\author{Carmen S\'{a}nchez Contreras}
\affiliation{Centro de Astrobiologia (CAB), CSIC-INTA. Postal address: ESAC, Camino Bajo del Castillo s/n, 28692, Villanueva de la Ca\~nada, Madrid, Spain}
\author{Michael E. Ressler}
\affiliation{Jet Propulsion Laboratory, California Institute of Technology, Pasadena, CA 91109, USA} 


\begin{abstract}
The planetary nebula (PN) NGC\,3132 is a striking example of the dramatic but poorly understood, mass-loss phenomena that $(1-8)$\,\ms~stars undergo during their death throes as they evolve into white dwarfs (WDs). From an analysis of JWST multiwavelength ($0.9-18$\,\micron) imaging of NGC\,3132, we report the discovery of an asymmetrical dust cloud around the WD central star (CS) of NGC\,3132, seen most prominently in the 18\,\micron~image, with a surface-brightness limited radial extent of $\gtrsim2{''}$. We show that the A2V star located $1\farcs7$ to CS's North-East (and 0.75\,kpc from Earth) is gravitationally-bound to the latter, by the detection of relative orbital angular motion of $0.24\arcdeg\pm0.045\arcdeg$ between these stars over $\sim$20\,yr. Using aperture photometry of the CS extracted from the JWST images, together with published optical photometry and an archival UV spectrum, we have constructed the spectral-energy distribution (SED) of the CS and its extended emission over the UV to mid-IR ($0.091-18$\,\micron) range. We find that fitting the SED of the CS and the radial intensity distributions at $7.7, 12.8$ and $18$\,\micron~with thermal emission from dust requires a cloud that extends to a radius of $\gtrsim$1785\,au, with a {\bf dust} mass of $\sim1.3\times10^{-2}$\,M${_\oplus}$, and grains that are 70\% silicate and 30\% amorphous carbon. We propose plausible origins of the dust cloud and an evolutionary scenario in which a system of three stars -- the CS, a close low-mass companion, and a more distant A2V star -- forms a stable hierarchical triple system on the main-sequence but becomes dynamically unstable later, resulting in the spectacular mass-ejections that form the current, multipolar PN.
\end{abstract}

\section{Introduction}\label{intro}
In our quest to identify the signatures of extraterrestrial life, the search for planets and planetary systems around stars other than our Sun has become one of the most exciting areas of current astrophysical research. Planets are found to be common around other solar-type stars, but the disks in which these are produced dissipate as the stars reach the main-sequence. A Spitzer 24\,\micron~survey of main-sequence A-type stars shows that up to $\sim$50\% of young ($\lesssim$30 Myr) stars have little or no 24\,\micron~excess emission from debris disks; and large debris-disk excesses decrease significantly at ages of $\sim$150 Myr, and much of the dust detected in these is likely produced (episodically) by large planetesimal collisions \citep{Rieke05}. The dust in these debris disks would have dissipated long before the stars evolve off the main sequence. Thus it is remarkable that once these stars reach the end of their life cycle, observational surveys reveal disks or equatorially-flattened disk-like or toroidal structures (e.g., \cite{Hillen2017,Sahai2007,Sahai2011}) -- i.e., both the birth and the death of stars represent similar highly aspherical states that sandwich the more spherical life of stars on the main sequence (MS).

The demise of most stars in the Universe that evolve in a Hubble time (i.e., in the 1--8\,\ms~range) is believed to occur as a result of heavy mass-loss (with rates up to $10^{-4}$\,\my) on the Asymptotic Giant Branch (AGB), when the stars are very luminous ($L\sim5000-10,000$\,\ls) and cool ($T_{eff} < 3000$\,K) (see e.g., review by \cite{Decin2021}). 
After mass-loss -- which may be via a quasi-steady wind over $\sim10^5$\,yr or rather sudden (if driven by a strong binary interaction, as e.g., in the Boomerang Nebula \citep{Sahai2017}) -- has depleted most of the stellar envelope, the stars evolve to higher temperatures through the post-AGB phase at almost constant luminosity, fading and becoming white dwarfs (WDs) at the ends of their lives. It is during these post-AGB and WD phases, that the renewed presence of puzzling disk-like structures around the central stars becomes obvious. How are such disks formed? Are there multiple mechanisms that can make disks during this phase? Can some of these be sufficiently dense (and long-lived), leading to a second phase of planet-formation?

Such disks can generally be divided into two broad observational classes by their sizes -- (Type 1) very small disks which are contained within the Roche limits of the central WD stars ($\lesssim0.01$\,AU), and (Type 2) large disks extending to radii of $\sim10-100$\,AU. The majority of the Type 2 disks are found around the central stars of PNe (CSPNe) (e.g., \cite{Bilikova12}) and around post-AGB stars (e.g., \cite{DeRuyter06}), whereas most Type 1 disks have been identified around old, naked WDs (i.e. those for which the PN shell is no longer visible).
Type 1 dust disks around WDs were first detected around G29-38 and GD 362 through their excess IR emission (confirmed spectroscopically to be continuum), and since their discovery, it has been commonly accepted that they originate from tidally disrupted planetesimals (\cite{Becklin05,Kilic05,Jura03}), especially since the chemical composition of these planetesimals is similar to rocky bodies in the inner regions of our Solar System. The required gravitational scattering of planetesimals towards the WD implies the presence of a planet, and recent discoveries provide indirect and direct evidence for these. The WD J091405.30+191412.25 is found to be accreting hydrogen, oxygen, and sulfur, material which likely comes from the deep atmospheric layers of an icy giant planet \citep{Gansicke19}. From TESS data, \cite{Vanderburg20} find a giant planet transiting the WD 1856+354 every 1.4 days. However, tidal shredding of a companion during a Common Envelope phase can also make such a disk\,\citep{Kuruwita16}.

The origin(s) of Type 2 disks is(are) more difficult to understand. Using Spitzer/MIPS imaging and IRS spectroscopy, \cite{Su2007} find evidence for the presence of a compact dust cloud around the central star of the Helix Nebula (WD 2226-210: $T_{eff}=110,000$\,K) with a temperature of 90--130 K, located 40--100\,AU from the central star, and inferred it to be a dusty disk. 

A more extensive search for 24\,\micron~excesses around a sample of CSPNe by (\cite{Bilikova12,Chu11}) revealed spectral-energy-distributions (SEDs) and spectra (for a few objects) with a variety of IR excess characteristics, implying the operation of a different mechanism than planetesimal destruction for producing the dusty disk responsible for the excess emission. 
This mechanism involves the presence of a binary companion -- binaries provide a source of angular momentum and free energy to form accretion discs. Several mechanisms for making disks in binary systems with a mass-losing companion have been identified (including Bondi-Hoyle-Eddington accretion, Roche-lobe and wind-Roche-lobe overflow, common envelope ejection, grazing envelope ejection).
Numerical simulations of binary systems where the companion is detached show that the latter can gravitationally capture a fraction of the mass outflow from an AGB star into a small disk (hereafter Type 1b, $\sim$1\,AU, \cite{Mastrodemos98}) -- too small to be representative of Type 2 disks. Thus, the large Type 2 disks, if resulting from binarity, likely require interaction with a close binary companion that results in the ejection of a substantial fraction of the AGB mass outflow being directed along directions near the orbital plane forming a disk-like structure.
But it has been observationally difficult to confirm the association of dust disks with binary CSPNe, which requires the detection of companions and disks around the same CSPNe. The spectral-type distribution for companions of WDs (which should be representative of CSPNe), peaks at spectral types M\,3-4 (\cite{Farihi05}).
Spectra or sensitive photometry in the $\sim0.6-5$\,\micron~range can reveal a companion -- e.g., an M3V companion would emit $\sim210 (D/0.5\,kpc)^2$\,$\mu$Jy at 5\,\micron~-- or provide upper limits.

In this paper, we report the detection of a dust cloud around the central star (CS) of the planetary nebula NGC\,3132 using data from the JWST's Early Release Observations (\cite{Pontoppidan2022}). This PN, also known as the Southern Ring, has been well-studied, with high-resolution optical images obtained with HST (e.g., \cite{Monteiro2000}), 2D spectroscopic imaging (\cite{Monreal-Ibero20}), and mapping of molecular-line emission  -- 2.1\,\micron~H$_2$ (\cite{Storey84}) and 1.3\,mm CO J=2--1 (\cite{Sahai1990}). These data reveal that NGC\,3132 belongs to the primary morphological class ``multipolar" (\cite{Sahai2011}) -- with a bright central elliptical shell structure surrounded by fainter structures aligned along different axes. Located at the center of the elliptical structure, the CS of NGC\,3132 is a hot white dwarf (discovered by \cite{Kohoutek77}) with $T_{eff}\sim105,000$\,K: \cite{Monreal-Ibero20}). North-East of the CS, at a separation\footnote{all separations are as measured on the sky-plane, unless noted otherwise} of $1\farcs696$, lies a companion (HD\,87892) that is a slightly evolved main-sequence star of spectral type A2V (\cite{Mendez78}).

We focus in this study on the CS and its immediate surroundings; description and analysis of the full PN morphology is not within the scope of this paper, but has been reported by \cite{DeMarco2022}.


\section{Observations}
We retrieved the pipeline-calibrated Level3 data on NGC\,3132 (Proposal 2733, PI=K.M. Pontoppidan) from the MAST archive. NGC\,3132 was imaged using filters  F090W, F187N, F212N, F356W, F405N, F470N with NIRCam (plate-scale $0\farcs031$ for $\lambda\lesssim2.1$\,\micron~and $0\farcs063$ for longer wavelengths) (\cite{Jakobsen2022}), and F770W, F1130W, F1280W, and F1800W with MIRI (plate-scale $0\farcs11$) (\cite{Bouchet2015}). JWST is diffraction-limited at wavelengths $\gtrsim$2\,\micron~with a PSF size (FWHM) of $0\farcs075$ at $\gtrsim$2\,\micron~(\cite{Rigby2022}). The images have been processed and calibrated using data processing/calibration software version nos. ``2022\_2a" and ``1.5.3". Each of the Level 3 images is a combination of 8 dithered images (hereafter ``sub-images"). For NIRCam data, 6/8 sub-images in each of the filters F090W, F187N and F202N include the CS in the FOV; for the remaining filters, the CS is within the FOV for all 8. For MIRI data, 6/8 sub-images in each of the filters include the CS in the FOV. We have examined each of these sub-images to look for artifacts that may affect the CS, and found that only one of these, for F090W, has 4 bad pixels at the location of the CS. Hence, for this filter only, the photometry described in the next section (\S\,\ref{obsres}) was carried out on each of the 5 good sub-images, and the reported value is a median average of these.

\section{Observational Results}\label{obsres}
The JWST images of NGC\,3132 reveal the well-known, extended multipolar morphology of this object, the hot central star that excites the nebular material, and the nearby A2V star. 
The images (in selected filters) of the CS and its immediate surroundings, are shown in Fig.\,\ref{cs-miri-nirc}. A small but clearly extended emission region is seen around the CS in the F1800W image. In addition, the emission is not circularly symmetric in its outer parts (i.e., at radial offsets $\gtrsim 0\farcs65$) -- the contours appear flattened on the side closest to the A2V star, thus giving the emission source an overall elongated shape (Fig.\,\ref{cs-f1800}a,b). Furthermore, there is a small ``tail" at the Northern side of the emission source, that bends towards the A2V star. Radial intensity cuts centered on the CS, averaged over 90\arcdeg~wedges pointing towards and away from the A2V star, clearly show a significant mismatch between the intensities at offsets $\gtrsim0\farcs7$ (Fig.\,\ref{cs-f1800}c). Although there is considerable structure in the foreground/background of the CS due to the PN itself, it is very unlikely that the above asymmetries are the result of a chance projection of a large extended nebular feature coinciding with the central dust clump, distorting its shape and producing such a specific tail shape curved towards the A2V star. Also, no hint of such an extended feature close to the compact dust clump is seen in the large-scale 18\,\micron~image of the nebula (Appendix).

A comparison of the CS's radial intensity with that of relatively-bright field stars in the field-of-view of these images, shows that at 18\,\micron, the width (FWHM) of the emitting region, $0\farcs96$, is significantly larger than that of the PSF ($0\farcs66$). 
The observed radial extent of the dust cloud is surface-brightness limited -- e.g., on the side away from the A2V star, it falls to an intensity equal to the standard deviation of the sky background, $\sim$4\,MJy\,sr$^{-1}$, at $r\sim 2\farcs5$.

The 12.8, 11.3 \& 7.7\,\micron~images also appear mildly extended, but to a progressively lesser extent; for shorter wavelengths, the CS appears to be point-like.

\subsection{PSF subtraction}
The CS images are affected by the presence of the bright A2V type star which spreads its flux through its PSF. 
We describe our 5-step (iterative) PSF subtraction process below:  

\begin{enumerate}
\item Generate a simulated PSF (``sPSF") using the PSF simulator tool ({\tt webbpsf}{\footnote{https://www.stsci.edu/jwst/science-planning/proposal-planning-toolbox/psf-simulation-tool}}), taking into account the spectral type of the source (A2V) and its offset from the center of the FOV in the NIRCam and MIRI images. The optics of the system are affected by additional effects (jitter, deformations,...) which are equivalent to an additional gaussian smoothing but are not accounted for in the simulator. We therefore applied a gaussian smoothing function to ``sPSF", whose FWHM was set iteratively (see step 3 below).
\item Remove a median averaged sky-background from the ``i2d" image. The value for the box size used for estimating a median background was set iteratively for each filter/camera.
\item Apply the gaussian smoothing function to the sPSF, and scale the sPSF peak (central) intensity to that of the A2V star when the latter was not saturated in the center (F1800W, F1280W, F1130W, F770W), or to the intensity in the wings when the A2V star is saturated in the center (F090W, F187N, F212N, F356W, F405N, F470N). The width of the gaussian was set iteratively for each filter/camera.
\item Compare intensity slices along the vertical and horizontal axis crossing through the center of the A2V star and the sPSF, and check the result of the subtraction of the current sPSF with the background-subtracted image. This allowed us to estimate the background subtraction performed in step 3, as well as re-adjust the gaussian spread applied to the PSF. Also, the registration of  the sPSF relative to that of A2V star was re-adjusted, especially important in the NIRCam images for which the central pixels covering the A2V star are saturated.
\item Iterate steps 2--4 (i.e. gaussian smoothing, registration, flux scaling) to bring the average flux in the region close to the A2V star as close to the median error in the sky background in this region.
\end{enumerate}

\begin{figure}[ht!]
\includegraphics[width=1.0\textwidth]{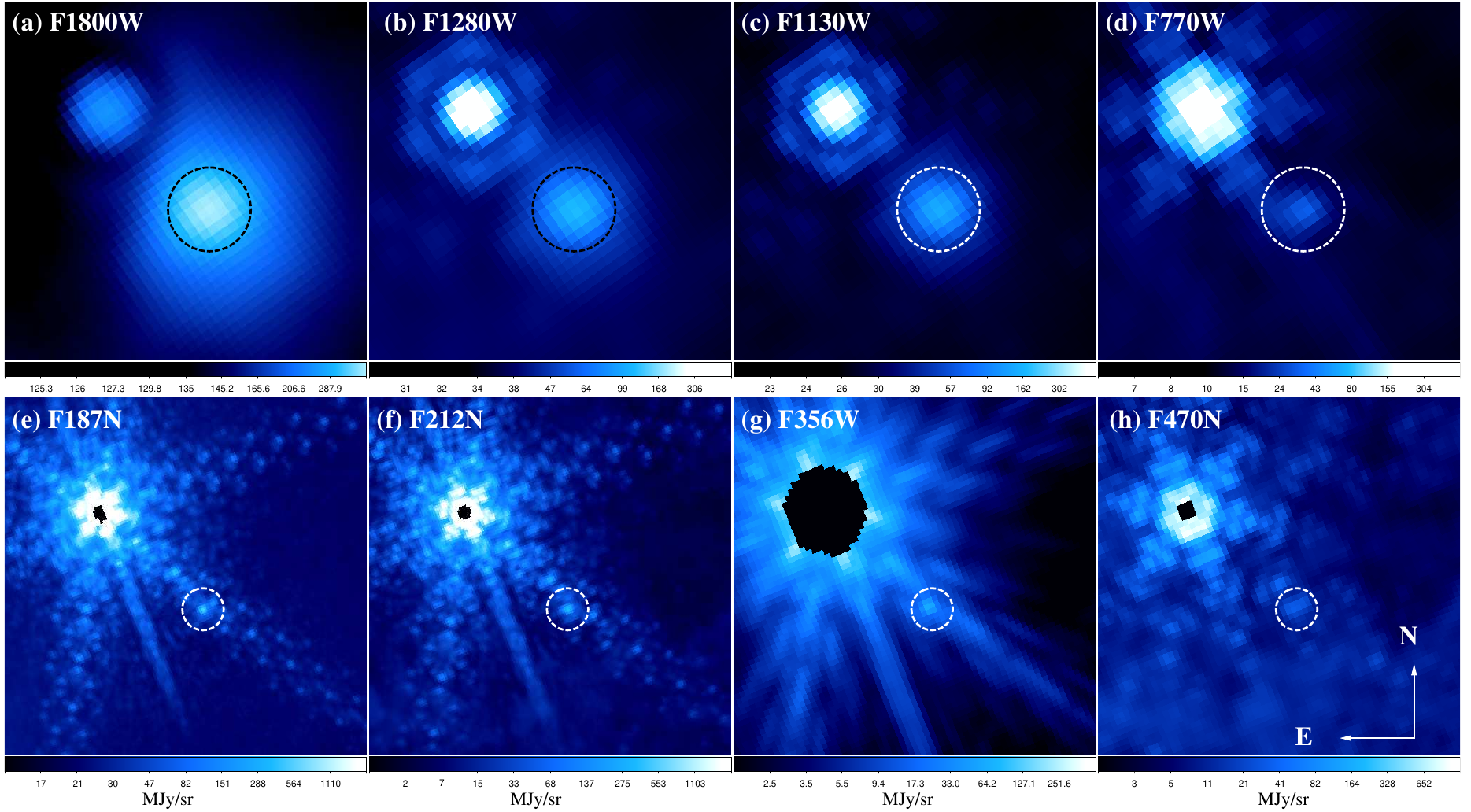}
\caption{Images of the central region of NGC\,3132, taken with MIRI (a) F1800W, (b) F1280W, (c), F1130W, (d) F770W), and NIRCam (e) F187N (f) F212N (g) F356W (h) F470N, shown using a logarithmic strech for the intensity (MJy/sr). Black (white) dashed circles of diameter $1\farcs0$ ($0\farcs5$) locate the central white-dwarf star (CS) in the MIRI (NIRC) images; the A2V companion is located 1\farcs7 to the NE of the CS.}
\label{cs-miri-nirc}
\end{figure}

\begin{figure}[ht!]
\includegraphics[width=1.0\textwidth]{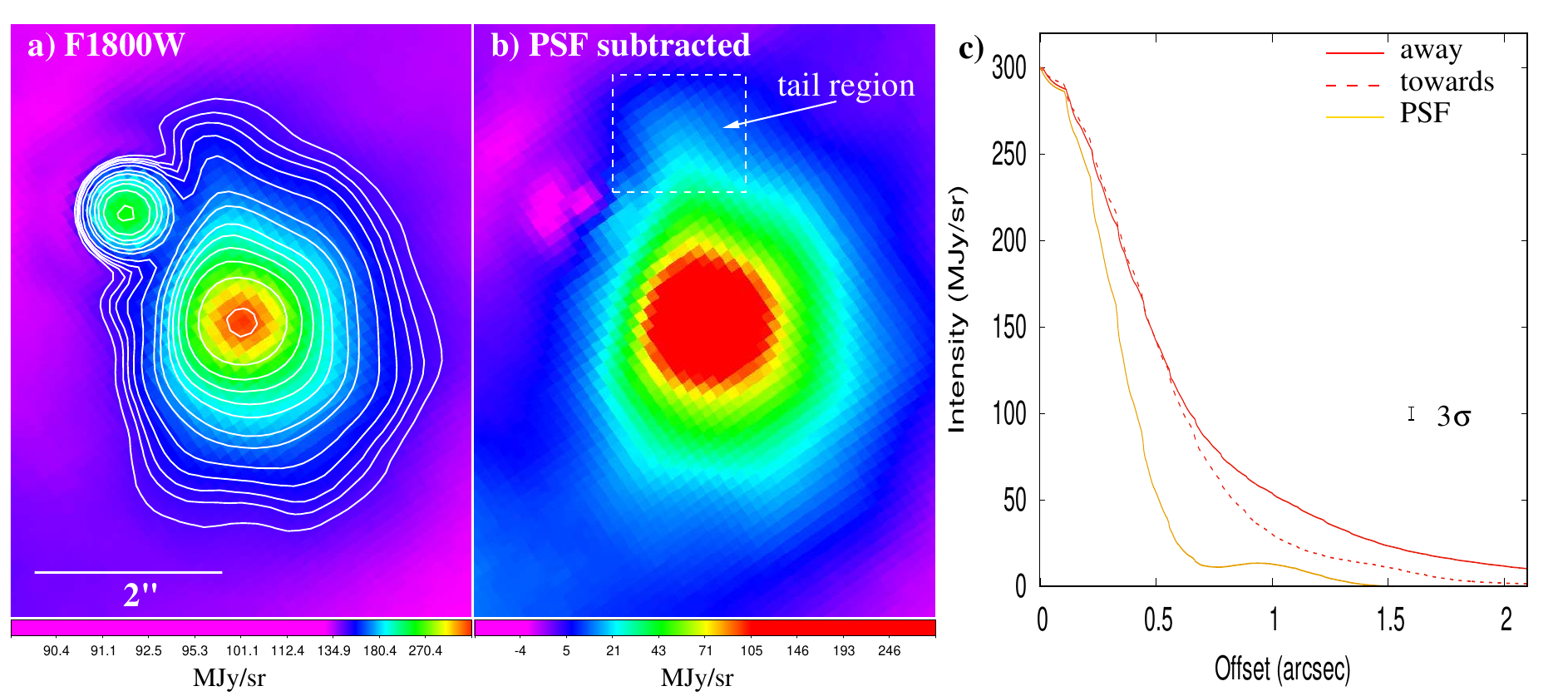}
\vskip 0.1in
\caption{Images of the dust cloud around the CS, taken with F1800W (a) pipeline image (logarithmic stretch), (b) PSF-sbtracted image (square-root strech). Contour levels are at 0.88, 0.50, 0.33, 0.25, 0.20, 0.167, 0.125, 0.10, 0.067, 0.05, 0.04, 0.033, 0.025 times the background-subtracted peak intensity (308\,My/sr) in panel $a$; the median background level is 135\,MJy/sr. Dashed box in panel $b$ demarcates the region of the ``tail" feature. (c)  Radial intensity cuts centered on the CS, averaged over 90\arcdeg~wedges pointing  away (solid red curve) and towards (dashed red curve) from the A2V star. Brown curve shows the PSF extracted using a field star.}
\label{cs-f1800}
\end{figure}

\begin{figure}[ht!]
\includegraphics[width=1.0\textwidth]{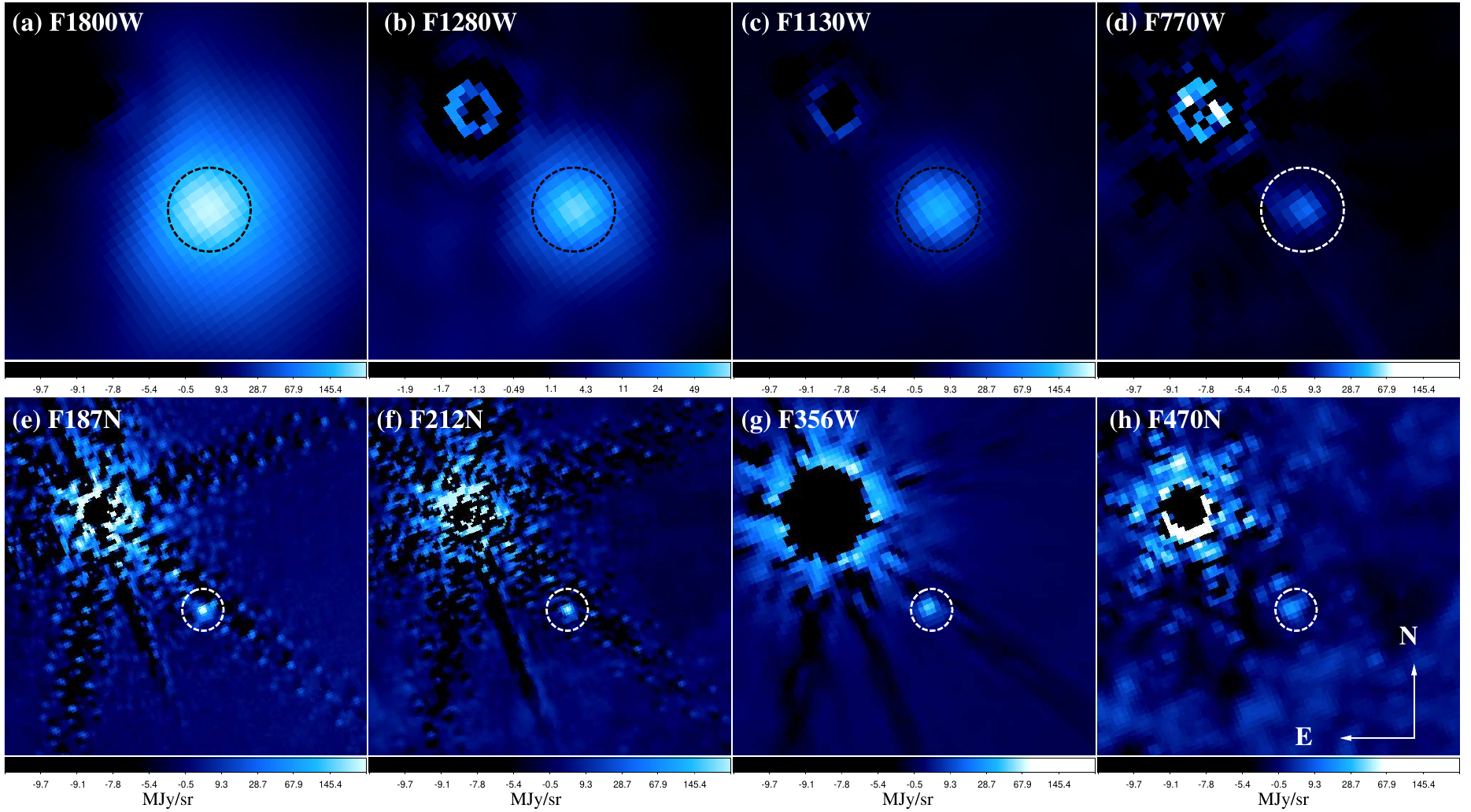}
\caption{As in Fig.\,\ref{cs-miri-nirc}, but with the A2V star subtracted.}
\label{cs-miri-nirc-psfsub}
\end{figure}

The background estimate and removal was performed by using the {\tt photutils}\footnote{https://photutils.readthedocs.io/en/stable/background.html} package (\cite{Bradley2022}). We estimated a 2D median background, by dividing the field of view into boxes and substituting the pixel intensities in each box with the median background for that box. This generated a low-resolution background which was subtracted from the image. The box size was estimated to remove as much background as possible but avoiding any source removal, including the PSF sidelobes. The box size, the width of the Gausian smoothing function, and the error in the PSF scale-factor\footnote{The actual value of the scale factor is unimportant, since it depends on the normalisation of the PSF, which is arbitrary} are given in Table\,\ref{tbl-psf}. The images of the CS and its immediate surroundings with the A2V star subtracted, for the same set of selected filters as in Fig.\,\ref{cs-miri-nirc}, are shown in Fig.\,\ref{cs-miri-nirc-psfsub}. Most of the negative residuals in these images lie relatively close to the sharp diffraction structures and most likely are a result of inherent limitations in the PSF simulator tool's ability to reproduce the observed PSF.





\subsection{Photometry}
We have carried out aperture photometry of the CS in the NIRCam and MIRI images as follows. Except for the F1800W image, we have used circular apertures with sizes adjusted to include as much of the light from the CS, and exclude as much light from the surrounding structured background, as possible. 

For F1800W, we have used multiple methods to extract the photometry. Using the pipeline Level 3 image, we use an elliptical aperture of size $1\farcs1\times1\farcs3$ that roughly matches the observed elongated shape of the source. Second, we extracted the average radial intensity over an angular wedge covering the position angle (PA) range $100\arcdeg-315\arcdeg$ that best excludes the region contaminated by the PSF of the A2V star; the background intensity was estimated from the cut intensity at large radii and subtracted. The resulting radial intensity was then integrated over circular apertures with varying outer radii to estimate the flux density as a function of aperture radius (Table \,\ref{tbl-flux-cs}). Radial intensity cuts have also been extracted from the F770W, F1130W and F1280W filter images in a similar manner as for F1800W.


The background appears to be dominated by nebular emission in the F1130W, F1280W and F1800W filters; for shorter wavelengths the PSF of the A2V star becomes increasingly dominant while the CS emission becomes progressively fainter. Aperture corrections have been determined using relatively bright field stars in the images, and applied to the aperture photometry of the CS (Table\,\ref{tbl-flux-cs}). In order to assess errors, for each filter, we used apertures of different diameters, each with its own aperture correction; the results shown in Table\,\ref{tbl-flux-cs} are those obtained with largest aperture that could be feasibly used. We have assigned conservative errors of $\pm$15\% to the NIRCam photometry, and $\pm10$\% to the MIRI photometry. Noting the 0.1 to 0.2 mag difference between pipeline and manually-determined photometry for the HST data at 0.44 and 0.56\,\micron~(Table\,\ref{tbl-flux-cs}), we have assumed $\pm$15\% errors for the published photometry at shorter wavelengths, i.e. $\lambda<0.9$\,\micron. The aperture photometry was also carried out for the PSF-subtracted images, and yielded consistent results.


We have verified that our photometry methodology is correct by measuring the F356W and F770W flux densities for a few field stars, and comparing these to those measured with Spitzer, IRAC\,1 and IRAC\,4 (Spitzer Science Center, 2021) -- we find that our photometry is in good agreement with the latter (Table\,\ref{tbl-flux-fs}), accounting for the possibility of temporal variability in these stars.

\subsection{UV Spectrum}
We downloaded a FUSE UV spectrum of NGC\,3132 that covers the $910-1190$\,\AA~wavelength range with a spectral resolving power $R\sim20,000$  (\cite{Moos2000,Sahnow2000}) from the Mikulski Archive for Space Telescopes, calibrated with the final version of the CalFUSE calibration pipeline software package, CalFUSE 3.2.3 (\cite{Dixon2007}). The spectrum was obtained as part of FUSE program ID P133 (PI: L. Bianchi), and an analysis of the same (reporting the discovery of Ge\,III\,$\lambda$\,1088.46 in NGC\,3132 and other PNe) was first presented in \cite{Sterling2002}. The spectra are of relatively modest S/N at the short wavelength end, and we have re-binned it to reduce the resolution by a factor 50. The spectrum was taken through the LWRS aperture, which has a size of $30{''}\times30{''}$, and therefore covers not only the CS, but also the A2V companion and a significant fraction of the ionized region. However, the continuum flux in the $910-1190$\,\AA~wavelength region is dominated by the CS, allowing us to use the FUSE spectrum to complete the SED of the CS to 0.091\,\micron~at the short wavelength end (see \S\,\ref{dustymod}).

\section{Analysis}

\subsection{A Central Binary and Orbital Motion}\label{cs-a2v-orbit}
We have investigated the relationship between the CS and the A2V star as follows. First, 
there is little doubt that these two stars are a physical pair \citep{DeMarco2022}, given the close agreement in their $Gaia$-based proper motions (A2V star: $PMra=-7.747\pm0.027$\,mas\,yr$^{-1}$, $PMdec=-0.125\pm0.031$\,mas\,yr$^{-1}$: CS: $PMra=-7.677\pm0.235$\,mas\,yr$^{-1}$, $PMdec=0.197\pm0.28$\,mas\,yr$^{-1}$) and radial velocities (A2V star:  $VLSR=-24.1\pm1.6$\,\kms, CS: $VLSR=-25\pm0.9$\,\kms). We have therefore inferred a distance to NGC\,3132 of 0.75\,kpc by inverting the parallax of the A2V star companion\footnote{no parallax is available for the CS} ($1.3198\pm0.0344$\,mas) listed in the $Gaia$ DR3 database \citep{Gaia2022}. Hence, the separation between the CS and its A2V companion and the CS is $1277\pm34$\,au. 

\begin{figure}[ht!]
\includegraphics[width=0.5\textwidth]{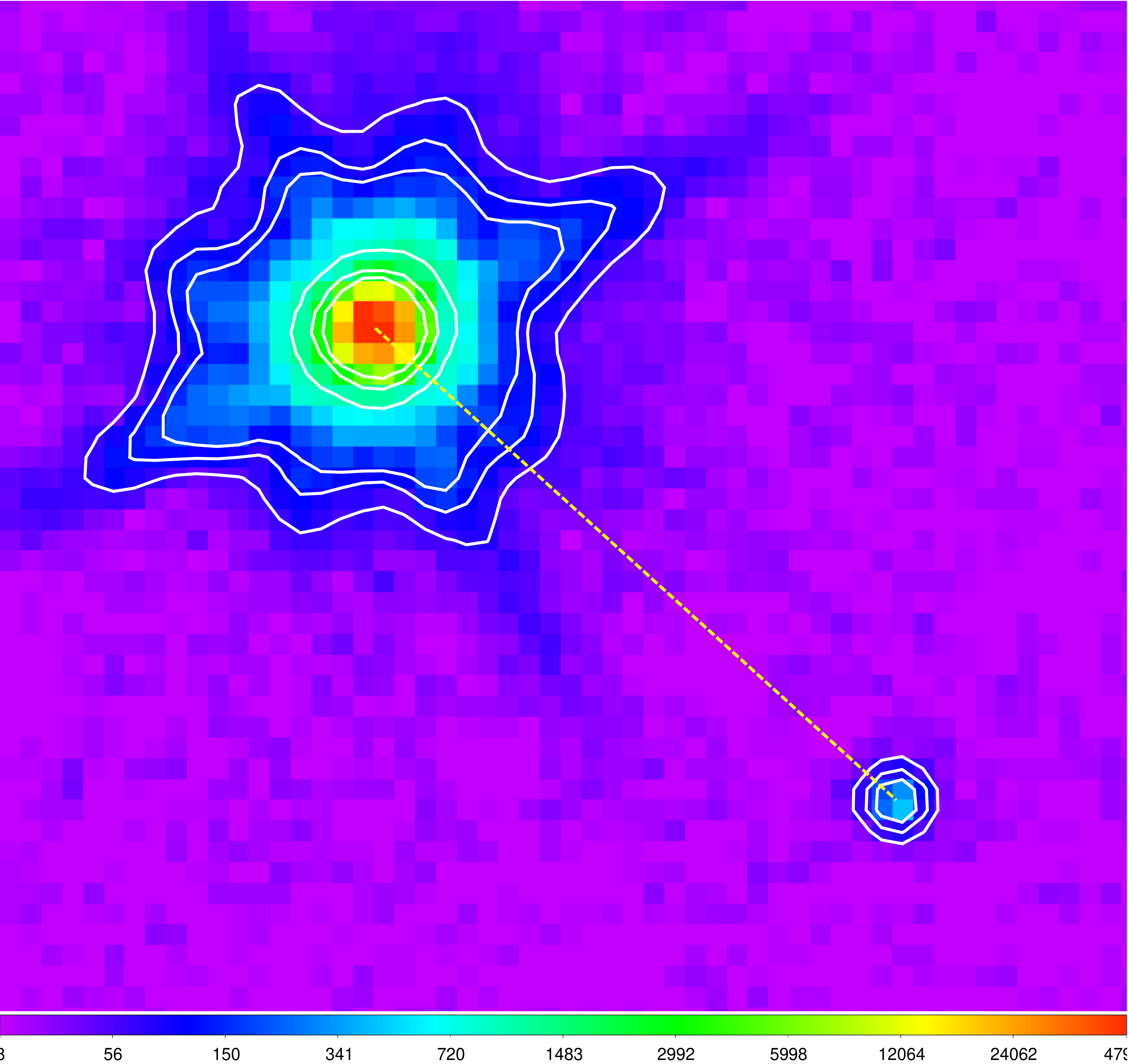}
\caption{HST 0.55\,\micron~(F555W) image of the CS and the A2V stars at the center of NGC\,3132. Dashed yellow line shows the vector separating the two stars, with length $1\farcs693$ and $PA=47.82\arcdeg$. North is up and East is to the left.}
\label{cs-hst}
\end{figure}

We have found direct evidence for the CS and the A2V star to be a gravitationally-bound binary system from the detection of orbital motion, over a time interval of $\sim20$\,yr between the epoch of the first high angular-resolution image of the CS-A2V pair with HST (Epoch 1), and the epoch for $Gaia$ measurements (Epoch 2) as follows. We used the Epoch 1 F555W image (from GO 06119, PI: H. Bond; image taken on 1995 Dec 04), obtained with the PC camera of WFPC2 (pixel-size of 0\farcs05), to determine the coordinates of the CS and the A2V star. Since the S/N is quite high ($>50$ for the CS, $>4000$ for the A2V), the uncertainty in the location of the CS is $\sim$1\,mas, and that of the A2V star much smaller. The position angle of the separation vector of the A2V,\,CS pair\footnote{measured from N towards E, using the CS as the origin},  in the HST image is, $PA(HST)=47.82\arcdeg\pm0.045\arcdeg$ (Fig.\,\ref{cs-hst}). The position angle of this vector during the $Gaia$ measurement epoch (Epoch 2: $\sim$JD$\sim$2457410), estimated from the DR3 coordinates of the CS and A2V stars, is $PA(gaia)=47.43\arcdeg$. 

The JWST data could not be used to determine the current orientation of the CS-A2V separation vector accurately enough because the locations of these stars cannot be measured with the required accuracy -- at the shortest wavelengths, where the telescope is diffraction limited and the PSF is small (e.g., $0\farcs075$ at $\sim2$\,\micron) the A2V star is badly saturated and the CS is very faint and its image is partly contaminated by the wings of the A2V star's PSF.

We have checked for any (minute) difference between the cardinal directions in the $Gaia$ DR3 reference frame ($Gaia$\,CRF3: \cite{gaia-ref-frame22}) and the HST image reference frame, $PA(gaia-HST)$, using the following methodology. We identified 4 field stars in the HST image (fs1, fs4, fs6, and fs8)\footnote{these are the 4/5 stars seen in the full FOV for which $Gaia$ DR3 data are available}, and computed the $PA$s of their separation vectors (Table\,\ref{tbl-pa-sep}).
Then, using the proper motions of these stars from $Gaia$ DR3, we have computed their expected locations in Epoch 1, and from these, the expected $PA$s of their separation vectors in Epoch 1  (Table\,\ref{tbl-pa-sep}). 
Because the separation vectors for these stars are quite large, the errors in the $PA$s are quite small ($\lesssim0\farcs01$). The average of the $PA(gaia-HST)$ values 
is $0.147\arcdeg\pm0.016$\arcdeg. Applying this correction to the value of $PA(HST)$ computed above, the corrected value of the $PA$ of the CS-A2V separation vector is,  $PA(HST,c)=PA(HST)-0.147\arcdeg=47.67\arcdeg$. 
Thus, the separation vector has rotated  by $0.24\arcdeg\pm0.045\arcdeg$ (clockwise) in 20.1\,yr (time interval between Epoch 1 and Epoch 2), in good agreement with the 
expected 0.20\arcdeg\,cos$(i)$/cos\,45\arcdeg~rotation for an orbital period of $P\sim25,500$\,yr, based on the measured separation of 1277\,au, assuming an intermediate angle between the orbital and sky planes of $i=45\arcdeg$ and masses of 0.7\,\ms~and 2.5\,\ms~for the CS and A2V stars.

The angular distance between the CS-A2V pair in Epoch 1 is $\epsilon(HST)=1\farcs693\pm0\farcs001$, compared to $1\farcs696$ in Epoch 2.
We have checked the ``plate-scale" of the HST image, using the same 4 fields stars and a similar methodology as above. We computed the angular distances between the star pairs used above for Epoch 1 using their current $Gaia$ DR3 coordinates and proper motions to compute their Epoch 1 coordinates, and compared these with the directly measured angular distances in the HST image  (Table\,\ref{tbl-pa-sep}): the average fractional difference for the distances between star pairs between Epoch 1 and 2 (Table\,\ref{tbl-pa-sep}) is $(3.92\pm1.35)\times10^{-5}$. This difference is too small to affect the value of $\epsilon(HST)$ derived above. Therefore, there is weak evidence for an increase in the angular distance between the CS-A2V pair of $2.7\pm1$\,mas from Epoch 1 to Epoch 2.


Although additional HST imaging is available at intermediate epochs, these images proved to be inadequate for determining the locations of the two stars with sufficient accuracy, because of one or more of the following factors: (i) the pixelsize was twice as large (e.g., $0\farcs1$ using the WFPC2 Wide-Field cameras), (ii) only emission-line images were available providing inadequate S/N to centroid the CS accurately, (iii) the core of the PSF of the A2V star appeared non-circular and distorted. 

\subsection{Dust Radiative Transfer Modeling}\label{dustymod}
\begin{figure}[ht!]
\includegraphics[width=0.4\textwidth]{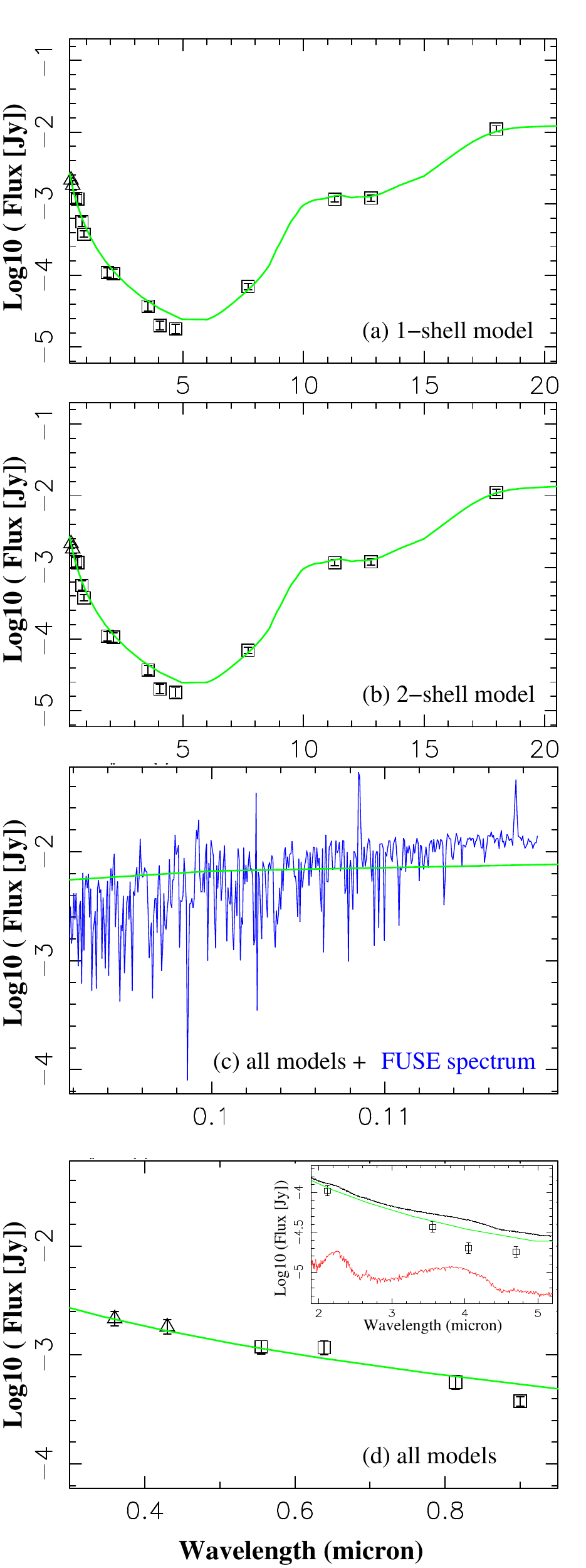}
\caption{Observed photometry (black symbols) and model spectral-energy-distributions (green curves) for the CS -- panels $a$ and $b$ respectively show the 1-shell and 2-shell models. Panels $c$ and $d$ show the model SED at short wavelengths where it is dominated by direct starlight (reddened and attenuated by nebular extinction) and is unaffected by the low-optical depth dust cloud, together with observed spectrum (panel $c$) or photometry (panel $d$). The inset in panel $d$ shows the model in panel $a$ or $b$ (green curve), an M6V dwarf companion (red curve), and the addition of the two (black curve). Error bars ($\pm15$\%) on the observed photometry are conservative estimates.}
\label{sed-obs-mod}
\end{figure}

We have used the MIRI and NIRCam photometry, together with the FUSE UV spectrum and published optical photometry, to construct the observed SED of the CS and the dust cloud around it over the UV to mid-IR ($0.091-18$\,\micron) region (Fig.\,\ref{sed-obs-mod}). 

The photometry is not co-eval; we assume that there are no significant variations in the CS's spectrum over the $\sim$4 decades spanned by epoch of the U-band photometry and that of the JWST observations.
Although the CS is included in the catalog of large amplitude variables from Gaia DR2 (Gaia DR2 5420219732233481472: \cite{Mowlavi2021}) is probably erroneous. \cite{Mowlavi2021} report a variability amplitude proxy of 
0.3566\,mag for the CS -- this proxy was computed from the uncertainty (keyword ``pho\_g\_mean\_mag\_error") in the G-band magnitude (keyword ``phot\_g\_mean\_mag") published in DR2. A comparison of the Gaia DR2 and DR3 G-band magnitudes  -- 15.216140 (mean magnitude based on 125 single epoch measurements) for DR2, and 16.105730 (mean magnitude based on 193 single epoch measurements\footnote{these include the 125 measurements used for the DR2 photometry}) for DR3 -- shows that the CS was almost one order of magnitude fainter in DR3 compared to DR2. However, since the DR3 data supersede the DR2 data, it is likely that the DR2 G-band magnitude and error are incorrect, and possibly result from incorrect removal of the contribution by the bright A2V star's PSF to the measured photometry of the CS.

The simplest explanation for the extended emission seen most clearly at 18\,\micron~is that it arises from thermal emission from warm dust. Although there are strong nebular emission lines that have been detected in NGC\,3132 in the wavelength bands covered by the filters listed above, it is very unlikely that such gas is present in the immediate vicinity of the CS. We have made a rough estimate of the width of the intrinsic 
radial intensity distribution at 18\,\micron~(FWHM$_{int}$) as follows. We approximated the observed radial intensity distribution as well as the PSF at 18\,\micron, using Gaussians of widths FWHM$_{obs}\sim1\farcs0$ and FWHM$_{psf}\sim0\farcs7$, and determined FWHM$_{int}\sim0\farcs7$ using ${\rm FWHM}_{obs}^2 = {\rm FWHM}_{int}^2 + {\rm FWHM}_{psf}^2$. This intrinsic width implies a radial extent of $\sim$260\,au at NGC\,3132's distance of 0.75\,kpc. Any gas this close to the star would be quite hot, $>10^4$\,K, and at the sound speed for such gas, 10\,\kms, would expand to radii $>10,000-20,000$\,au in (say) a typical post-AGB age of $>5,000-10,000$\,yr. Furthermore, the SED of NGC\,3132's CS dust emission is similar to that of the Helix, for which a spectrum of the CS shows no obvious line emission.



We have used the DUSTY dust radiative transfer code \citep{ivezic2012} to model the CS and its dust emission. Although it is plausible, depending on its origin, that the dust cloud is flattened or disk-like (see \S\,\ref{discuss}), since we have no direct information about this aspect, we have chosen to use 1D modeling. This is not a limitation, because (as shown below), the optical depth of this cloud is $\ll1$, even at relatively short wavelengths ($\lesssim0.001$ at $\lambda \gtrsim0.05$\,\micron), hence the results are not sensitive to the specific geometry (i.e., sphere or disk) of the cloud. Even if the dust cloud had a disk configuration, the radial optical depth near and in the equatorial plane would remain well below unity.

The main input parameters are (i) the dust temperature at the inner shell boundary ($T_{\mathrm d}$), (ii) the total radial optical depth at 0.55$\micron$~($\tau _V$), (iii) the shell density (iv) the grain-size distribution for a choice of grain composition, (v) the relative shell thickness ($Y$ = ratio of the shell's outer radius,  $R_{ou}$, to its inner radius, $R_{in}$), (vi) the spectrum of the central star. The shell density was assumed to be a single power-law exponent, $n$, of the power-law ($\rho_d(r)\propto r^{-n}$), or a combination of two power-laws (described below). We have computed the stellar spectrum using the T\"ubingen NLTE Model Atmosphere Package (TMAP) (\cite{Rauch2003,Werner2003,Werner2012}) for $T_{eff}=105,000$\,K -- as derived by \cite{Monreal-Ibero20} using photoionization modeling. We have accounted for nebular extinction by using a visual extinction of $A_V=0.31$ --  derived by \cite{Kohoutek77}) -- to attenuate and redden the model SED. 

We require the grains to be relatively cool ($\sim100$\,K at $r\sim1{''}$) in order for the average slope of the model spectrum from 7.7\,\micron~to 18\,\micron~to match the observed one (Fig.\,\ref{sed-obs-mod}). 
The shape of the model SED longwards of $\sim5$\,\micron, where dust emission is the dominant component, is most sensitive to the dust grain properties (composition, size distribution). We have investigated models with ``cold" (``Sil-Oc"), ``warm" silicate  grains (``Sil-Ow"), and amorphous C (amC) grains in DUSTY\footnote{DUSTY provides 6 in-built choices for grain composition}. The SED has a very distinctive shape in the wavelength region dominated by dust emission, i.e., longwards of $\sim5$\,\micron~-- there is a steep rise from $\sim7$\,\micron~to $\sim$11\,\micron, followed by a flat region between 11 and 13\,\micron, and then a rise towards longer wavelengths. While models with silicate grains produce the observed roughly flat shape of the SED in the 11-13\,\micron~region, pure amC-grain models produce a smooth rise with increasing wavelength, discrepant from the observed SED. Warm silicates produce modestly smaller ratios flux F(18\,\micron)/F(12.8\,\micron), F(18\,\micron)/F(11.3\,\micron) and F(18\,\micron)/F(7.7\,\micron) compared to cold silicates.
We used a modified version of the Mathis, Rumpl, Nordsieck (MRN) distribution function for grain radius $a$, $n(a) \propto$ a$^{-q}$ for $a_{\mathrm{min}}$ $\leq a \leq$ $a_{\mathrm{max}}$  (\cite{mathis1977}), with q=1.\footnote{The standard MRN parameters are: $q = 3.5$, $a_{\mathrm{min}}$ = 0.005$\micron$ and $a_{\mathrm{max}}$ = 0.25$\micron$.}  Models with a substantial fraction of large grains e.g., $a_{\mathrm max}\gtrsim1$\micron~produced radial intensity profiles that are significantly more compact than observed; in addition, the detailed shaped of the SED between 7.7 and 18\,\micron~(described above) cannot be reproduced -- large grains produce a dust emission spectrum in which there is a smooth rise with increasing wavelength. \cite{DeMarco2022} present a simple large-grain (size $\sim$100\,\micron) dust shell model, which suffers from these deficiencies.
Keeping the grain size distribution the same, but allowing the grains to be cooler by decreasing $T_{\mathrm d}$ results in a decrease in the 7.7\,\micron~ to 18\,\micron~flux ratio, and simultaneously makes the model emission more extended due to an increase in $R_{in}$.
Keeping the value of $T_{\mathrm d}$ fixed, an increase in the proportion of larger grains by increasing a(max) or decreasing $q$, makes the dust emission more compact. 
The model radial intensity distribution, especially at the longer wavelengths ($\sim11-18$\,\micron) is sensitive to (i) $T_d$, and (ii) the radial density gradient of the dust shell. Lower values of $T_d$ lead to smaller values of $R_{in}$, and therefore a more compact radial intensity distribution of the dust emission. Higher values of $n$ also result in more compact dust emission.

The above competing constraints allow us to provide reasonable constraints on the model parameters. 

The DUSTY code generates a model SED, normalized to the bolometric flux, $F_{bol}$. We find that $F_{bol}=(6.5\pm0.5)\times10^{-9}$\,\bfluxu~(implying a CS luminosity of 114\,\ls) by scaling the (reddened) model SED to match the observed SED (Fig.\,\ref{sed-obs-mod}a). The value of $F_{bol}$ is well-constrained (for a given choice of $T_{eff}$), independently of the specific dust model, because the model SED in the optical to near-IR wavelength range (i.e., $\sim0.1-3$\,\micron) (Fig.\,\ref{sed-obs-mod}d) is only affected by nebular extinction (which is modest), and not by the dust cloud which has a very low optical depth. While determining the best-fit model, we have ignored (i) photometry from the narrow band filters F187N, F212N, F405N, and F470N, as the data from these are most affected by the presence of nebular atomic or molecular hydrogen lines, and (ii) $Gaia$ DR3 photometry due to potential imperfect removal of the PSF of the A2V star. We did not convolve the SED with the filter responses, but we have checked that the difference between convolved and non-convolved model photometry is relatively small and well below difference between the best-fit model and the data.

In order to quantitatively distinguish between models, we have defined a `goodness-of-fit' measure for the SED modeling, $G(sed)$, as follows:
$G(sed)$ =$\sum \left(\frac{O_{\rm j}-M_{\rm j}}{\sigma_{\rm j}}\right)^{2}$, 
where, O$_{\rm j}$ ($\sigma_{\rm j}$) is the observed flux (error), M$_{\rm j}$ is the model flux, and the index j refers to different wavelengths. The summation was carried out only for $\lambda \ge 3.56$\,\micron~because at shorter wavelengths, the SED is dominated by the central star, and including the data for these would reduce the sensitivity of $G(sed)$ to the contribution of the dust emission. The `goodness-of-fit' for the radial intensity disributions, $G(Fx)$, where $Fx$ is F770W, F1280W, or F1800W, is:
$G(Fx)$ =$\sum [r_{\rm j}\,\delta\,r_{\rm j}\,(Oint_{\rm j}-Mint_{\rm j})/\sigma_{\rm j}]^2/\sum [r_{\rm j}\,\delta\,r_{\rm j}]^2$, 
where $Oint_{\rm j}$ ($\sigma_{\rm j}$) is the observed normalized intensity (error), $Mint_{\rm j}$ is the model normalized intensity, $r_{\rm j}$ ($\delta\,r_{\rm j}$) is the radial offset (width of annulus) at grid-point j,  and the index j refers to different radial offsets. The inclusion of $r_{\rm j}\,\delta\,r_{\rm j}$ in the formula for $G(Fx)$  means that we have used the difference of between (normalized) observed and model flux at each radial grid point for determining the goodness-of-fit. 
Better-fitting models have lower values of $G(sed)$ and $G(Fx)$.

\begin{figure}[ht!]
\includegraphics[width=0.5\textwidth]{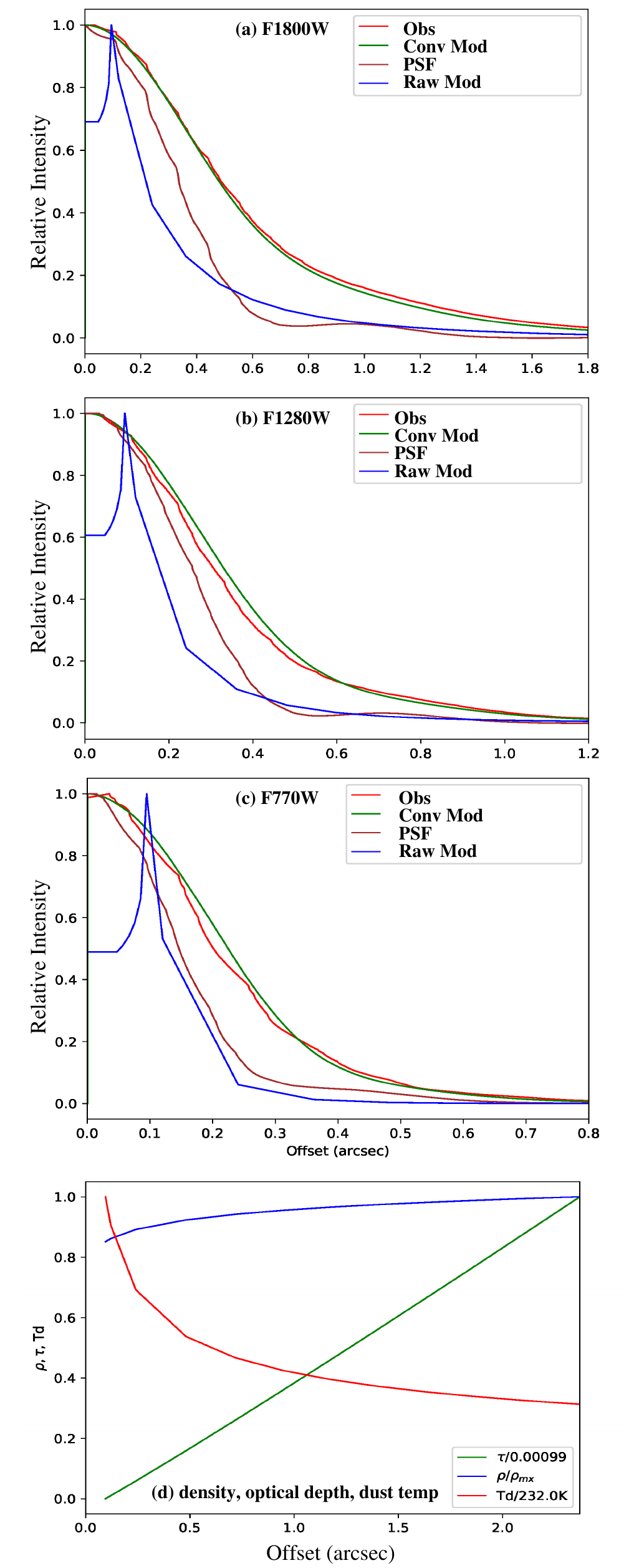}
\caption{Observed (azimuthally-averaged) and model radial intensity distributions of the CS (+dust) of NGC\,3132 as seen in the (a) F1800W, (b) F1280 and (c) F770W filters, together with (d) the radial distribution of density, visual optical depth, and dust temperature for the best-fit 1-shell model. All distributions have been normalized to their peak values. This 1-shell model uses a grain mixture consisting of 70\% warm-silicates and 30\% amorphous Carbon (amC). The model intensities (blue) have been convolved with the PSFs at 18, 12.8 and 7.7\,\micron, to generate the convolved models (green), for direct comparion with the observations (red). The sharp intensity peaks in the intrinsic model correspond to the inner radius of the dust shell. The F1280W and F770W intensities are shown over a smaller range of offsets compared to F1800W, because the observed radial intensity distributions for the former are much more narrow than for the latter.}
\label{radial-obs-mod-1sh}
\end{figure}

The above ``1-shell" models provide a reasonably good fit to the SED (Fig.\,\ref{sed-obs-mod}a) as well as the radial intensity in the F1800W, F1280W and F770W images  (Fig.\,\ref{radial-obs-mod-1sh}a,b,c). 
The grain composition needed to fit the specific shape of the SED, requires a mixture of 70\% cold silicate (Sil-cW) and 30\% amorphous carbon (amC) grains.
However, there are small but noticeable discrepancies  -- (i) the F1800W model radial intensity lies a little below the observed one for radial offsets $r\lesssim1{"}$, and (ii) the F1280W model radial intensity lies a little above the observed one for radial offsets $r\lesssim0\farcs65$.

Since the radial dust emission intensity is sensitive to the dust radial density distribution $\rho_d(r)$, we explore the possibility that a more complex density structure than the ones used for the 1-shell models can reduce the above discrepancy. We have investigated models in which $\rho_d(r)$ is described by a broken power-law, such that $\rho_d(r)\propto r^{-n1}$ for $1\le Y\le Y1$, and $\rho_d(r)\propto r^{-n2}$ for $Y1\le Y\le Y$ (hereafter ``2-shell" models). 

\begin{figure}[ht!]
\includegraphics[width=0.5\textwidth]{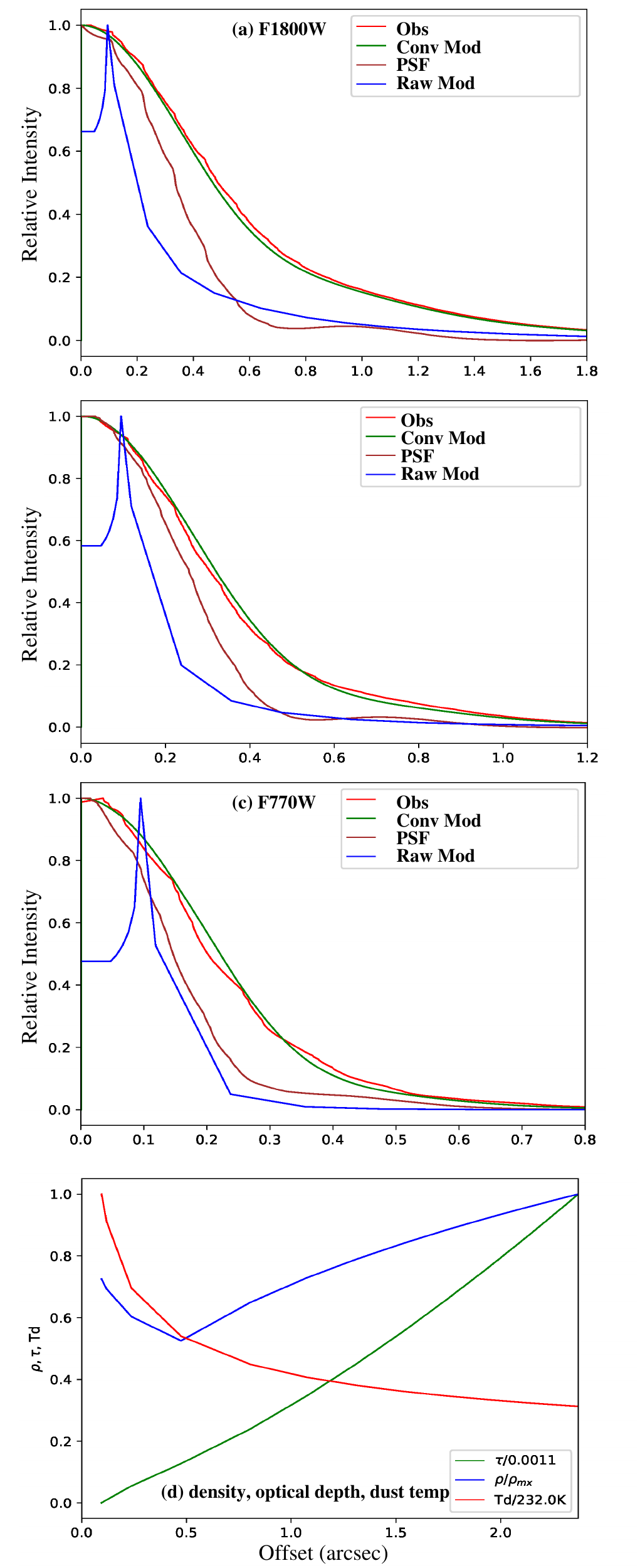}
\caption{As in Fig.\,\ref{radial-obs-mod-1sh}, but for a 2-shell model.}
\label{radial-obs-mod-2sh}
\end{figure}

We find that a 2-shell model with $n1=0.2$, $n2=-0.4$, and $Y1=5$ provides the best fit (Fig.\,\ref{radial-obs-mod-2sh}a,b,c). For this model, $G(sed)$ is a factor 0.75 times and $G(Fx)$ is a factor 1.2, 0.96, and 0.41 times (for $Fx$ equal to F770W, F1280W, and F1800W, respectively), of that for the 1-shell model.  While comparing the $G(Fx)$ values for these models, we give the most weight to $G(F1800W)$  and the least wight to $G(F770W)$ since the dust distribution is most (least) resolved at $18\,\micron$ ($7.7\,\micron$). Thus the 2-shell model provides an overall better fit to both the SED and the radial instensity distribution.


The density structure in the 2-shell model suggests that the dust cloud around the CS has two shells with a low density region in between.  The presence of double shells is not uncommon for the central stars of PNe that show dust emission -- \cite{Bilikova12} find the presence of double shells in $\gtrsim$50\% of the CSs with IR-excesses due to dust. 

We have therefore also investigated models with a different variant of the 2-shell density structure -- in this class of models (hereafter ``gap" models), we have a geometrically thin shell close to the star, separated by a density gap from the extended shell. In these models, we varied $T_{\mathrm d}$, the ratio of the density in the inner shell (assumed constant) to the density of the outer shell at the outer edge of the gap, the radius and width of the gap, and the power-law density exponent for the outer shell. We find that although the best ``gap" model
can fit the SED and the F1800W and F1280W radial intensities as well as the 1-shell and 2-shell models, the F770W radial intensity distribution is signifiantly narrower than observed. Hence, we do not discuss these models further.

We have derived the mass of the dust shell as follows, since DUSTY does not provide a direct measure of the shell dust mass.
For objects obeying a $r^{-n}$ density distribution, the dust mass in the circumstellar component is given by (for $n\ne1,3$): 
\begin{equation} M_{d} = 4\pi\,[(n-1)/(3-n)]\,y(Y)\,R^{2}_{in} (\tau_{18}/\kappa_{18})
\label{eqn-massd}
\end{equation}
where $y(Y)=(Y^{3-n}-1)/(1-Y^{1-n})$, and $\tau_{18}$ and $\kappa_{18}$ are, respectively, the radial optical depth of the shell and the dust mass absorption coefficient, at 18\,$\micron$. We assume $\kappa_{18}=10^3$\,cm$^{2}$g$^{-1}$, based on the dust properties for silicate dust tabulated by \cite{Ossenkopf1992}. For the `2-shell" models, we apply Eqn.\,\ref{eqn-massd} piecewise to the inner and outer shell. The input parameters and output properties for the best-fit models are given in Table\,\ref{mod-tbl}. The total dust mass is about $1.45\times10^{-2}$\,M${_\oplus}$. In the 2-shell model, the mass in the inner shell is a small percentage (0.5\%) of the total. For comparison, \cite{DeMarco2022} derive a dust mass of $\sim5\times10^{-2}$\,M${_\oplus}$ from their simple model.

For both the 1-shell and 2-shell model classes, models covering an appropriately large input parameter space were computed, allowing us to estimate rough uncertainties in the derived output parameters for a fixed value of $T_{eff}$. The most accurately determined parameter is $T_d$ (uncertainty $\lesssim5$\%), followed by $R_{in}$ (uncertainty $\lesssim10$\%). The uncertainty in the value of q is about 50\%; the maximum grain radius may be as high as $\sim0.5$\,\micron. The optical depth has an uncertainty of $\sim$20\%. The dust masses have an uncertainty of $\sim$50\%.


The temperature of the CS also affects the model results. 
At the relatively high effective temperature estimated for the CS, the flux at $\lambda\gtrsim0.4$\,\micron~follows the Rayleigh-Jeans approximation, i.e., $F(\nu,T)\propto(2\,k\,T_{eff}/\lambda^2)$, and $F_{bol}=F(\nu,T)\,\lambda^2\,(\sigma\,T_{eff}^3)/(2\,\pi\,k)$ where $k$ and $\sigma$ are the Boltzmann and Stefan-Boltzmann constants.
Hence the slope of the observed SED at short wavelengths (UV to near-IR), where the stellar emission dominates, does not provide a constraint on $T_{eff}$. Since $F_{bol}$ depends on $T_{eff}^3$ for a given value of the flux $F(\nu,T)$ and the latter is constrained by the observed photometry, variations in $T_{eff}$ lead to corresponding variations in the CS's inferred luminosity, $L=114\,\ls\,(T_{eff}/105,000K)^3$. Since a larger value of $L$ increases $R_{in}$ for a given value of $T_d$, making the model dust emission more extended than observed, models with larger $L$ require higher values of $T_d$. For example, if $T_{eff}=110,000$\,K, then  good 1-shell model fits (i.e., similar to the one shown in Fig.\,\ref{radial-obs-mod-1sh}) are obtained with $F_{bol}=(110/105)^3\,(6.5\times10^{-9})=7.5\times10^{-9}$\,\bfluxu, and $T_d\sim255-260$\,K, $\tau _V=8.6\times10^{-4}$, and mass $\sim(0.78-0.72)\times10^{-2}$\,M${_\oplus}$.

\section{Discussion}\label{discuss}
The dust cloud around the central star of NGC\,3132 is probably not the result of current mass loss, since such phenomena are not expected in a WD surrounded by a relatively old planetary nebula. It is more likely a stable disk in Keplerian or quasi-Keplerian rotation. As mentioned earlier (\S\,\ref{intro}), disks have been found to be associated with quite evolved WDs, often too old to show detectable PNe around them (e.g.\cite{Tokunaga1990, Manser2020}). These disks are very small, occupying just a few au, and have been proposed to be the result of disruption of planets or planetoids from a former planetary system.  Central stars of relatively young PNe tend to show disks with smaller radii, inferred to typically $\sim$ 50 au (\cite{Bilikova12,Su2007}). Many post-AGB stars (thus less evolved than old WDs) also show disks (e.g. \cite{DeRuyter06, Bujarrabal2013,Bujarrabal2016}), but in this case their extents are much larger, \gsim\ 1000 au. In this case, the rotating disks are systematically associated with the presence of tight binary systems and probably consist of material ejected during the previous AGB phase, which gains angular momentum from interaction with the system and forms rotating circumbinary disks.

We now discuss possible origins of the dust cloud around NGC\,3132's CS. 
\subsection{An Oort Cloud}
The large radial extent ($>1500$\,au) of the dust cloud in NGC\,3132,  makes it unlikely that it is formed from material extracted from a planetary system, or a Kuiper Belt/debris-disk analog -- the typical radius of the latter is about $\sim50-150$\,au. However, an Oort cloud analog might explain the origin of this dust cloud. The Oort cloud in our Solar system is thought to occupy a radius between 2,000 and 5,000\,au and may extend as far as 50,000\,au from the Sun. 

The dust cloud around the CS of NGC\,7293 may also be extended -- \cite{Su2007} find that the 24\,\micron~emission from this cloud, after background subtraction, appeared slightly resolved with a FWHM of $\sim9{''}$, 1.5 times that of a true point source -- although, as stated by \cite{Su2007}, this could be due to an imperfect subtraction of the background, it may also be real (as in the case of NGC\,3132.)

We now estimate the survival probability of Oort's rocky bodies after collisions with the gas and dust ejected by the CS while it was on the AGB or RGB. For a cometary nucleus with a typical radius of $\sim$1\,km, located at a distance of $\gtrsim2000$\,au, only about $0.56\times10^{10}$\,g of circumstellar ejecta will hit the nucleus assuming a total $\sim$1\,\ms~ejected via mass-loss, much smaller than the typical mass of the comet nucleus ($>10^{16}$\,g). So the cometary nuclei in the Oort cloud should easily survive the impact of the CS ejecta. Assuming that the impacting gas is expanding at 15\,\kms, and all of its linear momentum is transmitted to the cometary nucleus,  the latter will gain a velocity of 0.85\,cm\,s$^{-1}$, significantly smaller than the escape velocity at 2000\,au ($\sim1$\,\kms), implying that an Oort cloud will easily survive the AGB mass-loss phase. Thus, the Oort cloud, with an estimated mass of $\sim$1.9\,\me (\cite{Weissman1983}), can easily supply the mass of dust observed around the CS.

However, in this scenario, we require a mechanism to transport the dust inwards, since the inner radius of the dust cloud in our model is $\sim70$\,au. We think it is possible that the A2V companion provides sufficient perturbation to the orbits of the cometary bodies in the Oort cloud, causing them to collide and/or increase their eccentricity sufficiently so as to reach the inner regions of the dust cloud. 

\subsection{Interaction with a Binary Companion}\label{binaryaction}
As mentioned earlier, the central star of NGC\,3132 has a detached A2V sp.\,type companion (\S\,\ref{cs-a2v-orbit}). A closer companion has not been detected but cannot be discarded (we discuss this possibility in more detail below) -- the highest mass unresolved main-sequence companion that can be present and remain undetected is of lower mass than that of a M-dwarf with spectral type M6V (with L$\sim7\times10^{-4}$\,\ls~and $T_{eff}\sim2500$\,K, e.g., \cite{Cifuentes2020}), i.e., $\lesssim0.1$\,\ms. Including the theoretical spectrum of such a companion -- BT-NextGen (AGSS2009) model with $T_{eff}=2500$\,K, log\,$g$\,(cm\,s$^{-2}$)=5, extracted from http://svo2.cab.inta-csic.es/theory/newov2/index.php, \cite{Allard2011} -- increases the model SED flux in the F356W filter by 30\%, well above the observed value (see inset of Fig.\,\ref{sed-obs-mod}d).  \cite{DeMarco2022} propose that the CS has at least 2, and maybe even 3 close companions, in order to explain the origin of the dust cloud and the morphological structure of the extended PN.

From angular momentum balance considerations alone, the dust cloud may be a result of interaction with either such a close companion or the detached A2V companion. The angular momentum of the cloud, assuming it have a disk geometry and in Keplerian rotation around a $\sim0.65$\,\ms~CS, is low ($\lesssim0.5\times10^4$\,\ms\,km$^{2}$\,s$^{-1}$), orders of magnitude less than that of a compact binary system even if the secondary is just a big planet -- e.g., the angular momentum of a Jupiter-mass planet in a 1\,au orbit around the CS is $3.54\times10^{6}$\,\ms\,km$^{2}$\,s$^{-1}$. The A2V star's angular momentum, $\sim1.5\times10^{11}$\,\ms\,km$^{2}$\,s$^{-1}$, is also much larger than that of the CS's dust cloud, assuming the latter has a disk geometry.

From the point of view of the momentum transfer mechanism, however, the situation is very different for the above two scenarios. The currently adopted wind Roche-Lobe Overflow (wRLOF) mechanism of hydrodynamical interaction between a companion and a stellar wind in the red giant phase predicts strong effects when the separation of both stars is small, typically under 50--100 au (\cite{Mohamed2012,Chen2017,Kim2019}). The circumstellar gas at such distances is still being accelerated and shows a relatively low velocity, which is a basic ingredient for allowing strong interaction effects, including the formation of rotating circumbinary disks and symbiotic phenomena (e.g., \cite{SanchezContreras2022}).  Rotation in inner suborbital regions can also be induced in those cases of strong interaction (e.g. \cite{Bustamante2020}). However, when the distance between the two stars is significantly larger, the circumstellar expansion velocity is expected to be large and models predict that the gas passes by the companion with a minor interaction; the main effect is then the formation of spiral arcs due to the oscillation of the mass-ejecting primary. Formation of rotating disks is therefore not expected in detached binary systems. Hence, if the dust cloud is a disk resulting from binary interaction, we require that the CS has (or had in the past) another companion that is (was) much closer than the A2V star, and that this close companion (hereafter ``Comp$_{c}$") underwent a strong gravitational interaction with the CS.

The A2V star may have played an active role in inducing the strong gravitational interaction of Comp$_{c}$ with the CS. In this scenario, three stars -- the CS, Comp$_{c}$ (say at a separation of $\lesssim5$\,au), and the A2V star (in a more distant orbit, say at $\sim$400\,au) -- formed a stable hierarchical triple system while these stars were on the main-sequence. Such systems can become dynamically active on much longer time-scales due to the  ``Eccentric Kozai-Lidov" (EKL) mechanism causing the inner binary to undergo large-amplitude eccentricity and inclination oscillations (\cite{Kozai1962,Lidov1962,Naoz2016}). The oscillations tend to drive the inner binary to have very small pericenter distances and even to merge (e.g., \cite{Prodan2015,Stephan2018}).
\cite{Salas2019} have simulated such a triple system with a 2.2\ms~primary, a close companion (with a range of masses $\lesssim$0.9\,\ms) and a tertiary star with a range of masses M$\lesssim$0.9\,\ms) -- they find that in 37\% of their simulations, the tight binary merges. Although in the case of NGC\,3132, the tertiary star is more massive ($>$2.5\,\ms), \cite{Salas2019}'s results indicate that there is a significant possibility of the inner binary to merge. Such a merger would lead to a common envelope ejection of most of the stellar envelope of the primary star -- the multipolar morphology observed in NGC\,3132 would then be a result of interaction of multipolar collimated outflows with the ejecta, as appears to be the case in the Boomerang Nebula (\cite{Sahai2017}). The Boomerang is a pre-planetary nebula which has most likely resulted from CEE while the central star was still on the RGB and therefore in a much earlier post-AGB phase than NGC\,3132. Since the primary must be initially more massive than the A2V star, say $\sim2.8$\,\ms, simple conservation of angular momentum indicates that the tertiary's orbit, if initially equal to $\sim$400\,au, would expand to a semi-major axis of $\sim$1200\,au after the merger, bringing it to its current observed location.

\subsection{The A2V Companion's Effects on the Dust Cloud around the CS}
The JWST images show that the radial extent of the dust cloud is almost identical to the orbital radius of the CS-A2V detached binary system. In addition, there is a flattening of the 18\,\micron~intensity contours defining the shape of the dust cloud on the side facing the A2V star. These two features suggest a physical interaction between the dust cloud formation and the A2V companion; we now discuss several physical mechanisms for such an interaction. 
The first is that the A2V star preferentially illuminates the dust closest to it, making it relatively hotter, and therefore produce an extra brightening of the dust emission, relative to the diametrically-opposed side of the dust cloud, contrary to what is observed.

Next, we consider the effect of radiation pressure and a possible wind from the A2V star on the dust cloud around NGC\,3132's CS. The ratio of the stellar wind force ($F_{sw}$) to the gravitational force of star ($F_{gr}$) on a dust grain, is given by:
\begin{equation} 
\beta_{sw}=3 \mdot_{sw} V_{sw} C_D / (32\,\pi G M_{*} \rho_g a_g),
\end{equation}
where $M_{*}$, $\mdot_{sw}$, and $V_{sw}$, and are the mass, mass-loss rate and outflow velocity for the A2V star, $G$ is the gravitational constant, $\rho_g$ and $a_g$ are the dust grain material density and radius, $r$ is the radial distance from the A2V star, $C_D$ is a coefficient $\sim2$ (e.g., Eqn. 28 in \cite{Augereau2006}). Taking $\mdot_{sw}=10^{-10}$\,\my~(e.g., \cite{Lanz1992}), $V_{sw}=300$\,\kms, $\rho_g=3$\,\densunit, $a_g=0.25$\,\micron, $M_{*}=2.5$\,\ms, we get $\beta_{sw}=0.45$.
The ratio of the radiation pressure force ($F_{rd}$) to the gravitational force of star ($F_{gr}$) on a dust grain, $\beta_{rd}$, has been computed by \cite{Lamy1997} for stars of various spectral types and luminosity classes, two of which bracket the A2V star -- these are $\alpha$Aql (A5\,IV-V) and $\alpha$CMi (F6\,IV-V). This study shows that, for silicate grains of radii in the range 0.1--1\,\micron, $\beta_{rd}\sim1-3$, i.e., significantly greater than the value of $\beta_{sw}$ derived above. Thus, radiation pressure is much more effective in pushing the grains away from the A2V star than the A2V star's wind. Adopting an intermediate value, $\beta_{rd}=2$, the acceleration due to radiation pressure at a radial offset that is (say) halfway between the A2V star and the CS, i.e. at $r=0\farcs85$, is $\sim7.2\times10^{-6}$\,cm\,s$^{-2}$ (and varies as $r^{-2}$). Thus, within about 1000\,yr, radiation pressure will move the grains that are located (say) halfway between the A2V star and the CS, roughly 100\,au or $0\farcs13$ towards the CS. This pressure would push the outer regions of the dust cloud on the side facing the A2V star towards the CS, providing a plausible explanation for the observed flattening of the 18\,\micron~intensity contours defining the shape of the dust cloud on the side facing the A2V star. But this process would also lead to an increase in the column density near this edge, which would result in a brightening of the dust emission compared to the outer regions of the dust cloud that are located on the side facing away from the A2V star, contrary to what is observed.

A plausible explanation of why the dust cloud closer to the A2V star does not show enhanced emission, is that the wind from the A2V star also destroys some fraction of dust grains via sputtering. For example, e.g., \cite{Gray2004} have studied sputtering as a function of impact energy of hydrogen nuclei, and their Fig. 2 shows a substantial sputtering yield for impact energies of $\sim$1 keV (H and He nuclei moving at 300\,\kms~have energies corresponding to 0.47 and 1.9 keV).

Given the clockwise rotation that we find for the A2V star around the CS, the grains in the dust cloud must also be rotating clockwise (set by the sign of the global angular momentum of the primordial dense core in which these stars were formed). Hence, the dust grains approach the A2 star from the south-east as a result of their orbital motion. When they are far from the separation vector, the dominating force is the gravitational attraction to the CS, but that force becomes progressively less as the grains  get closer to the A2V star. The ``tail" seen in the 18\,\micron~image in the north/north-west direction is thus a signature of grains trapped by the A2V star (as they pass from a dynamical regime dominated by gravitation attraction toward the CS to one that is dominated by attraction toward the A2V star) or perhaps escaping the CS-A2V system (because they pass a region with a very weak net gravitational force).

\subsection{The Origin of the Density Discontinuity in the Dust Cloud}
The discontinuity in the 2-shell dust model suggests the presence of a radial gap in the dust cloud density, at a radius of $\sim$350\,au. A plausible origin of this gap is the presence of a giant planet or brown dwarf with an initial orbital radius of $\sim10$\,au; following the mass-loss from the central star as described above (\S\,\ref{binaryaction}), the orbital radius (as in the case of the A2V companion) would increase to $\sim$350\,au. This giant planet/ brown dwarf could then open up a fairly wide gap in the disk -- e.g., \cite{Crida2006} have computed gap density profiles using semi-analytic calculations and numerical simulations for a Jupiter mass planet in a disk, and find that the widths are comparable or larger than the orbital radius for low viscosities. The migration of the giant planet/ brown dwarf radially outwards could also contribute to further broadening the gap. Multiple planets would produce larger gaps but with shallower depths (e.g., see Fig.\,1 of \cite{Duffell2015}). 

\section{Conclusions}
We have analysed new imaging data of the central star (CS) of NGC\,3132 obtained using its NIRCam and MIRI instruments onboard JWST, through a set of filters spanning  the $0.9-18$\,\micron~range. Our main findings are as follows:
\begin{enumerate}
\item The CS is located at an angular distance of $1\farcs696$ from a bright A2V star to its north-east. We find that these stars form a wide gravitationally-bound system, separated by 1277\,au and located at a distance of 0.75\,kpc from Earth. The proper motions and radial velocities of these stars are consistent within uncertainties. In addition, we detect relative orbital motion between the two stars over a $20$\,yr period, which is consistent with the expected value from the 25,500\,yr period of the binary system estimated from the current separation and assuming an intermediate inclination angle of 45\arcdeg~between the orbital and sky-planes.

\item The A2V star outshines the CS at all but the longest wavelength of 18\,\micron. Using PSF subtraction of the A2V star, we find that CS is clearly seen in the JWST images even at the shortest wavelength 0.9\,\micron. The CS is surrounded by extended emission, seen directly at $18$\,\micron. Radial intensity cuts show that the emission is extended in the $7.7-12.8$\,\micron~range as well. This emission, which is surface-brightness limited and is somewhat asymmetrical in its outer regions, extends to a radius $\gtrsim$1600\,au and most likely results from thermal dust emission.

\item We have carried out aperture photometry of the CS in the JWST images. Using these data, together with an archival UV spectrum and published optical photometry, we have constructed the spectral-energy distribution (SED) of the CS and its extended emission over the UV to mid-IR ($0.091-18$\,\micron) range. The SED has a very distinctive shape in the $\sim5-18$\,\micron~region.

\item Using dust radiative modeling, we have fitted the SED of the CS and the radial intensity distributions at $7.7, 12.8$ and $18$\,\micron~with a dust cloud that extends to a radius of $\gtrsim$1785\,au. Models with a dust composition of 70\% silicate and 30\% amorphous carbon, and a modified MRN grain-size distribution that increases the proportion of larger grains, i.e., one in which the number of grains with radius $a$, varies as $n(a) \propto$ a$^{-q}$ for $a=0.005-0.25$\,\micron, with q=1, provide the best fit to the specific shape of the SED in the $5-18$\,\micron~range. The radial dust optical depth of this cloud is $\sim10^{-4}$ at 0.55\,\micron; and the dust temperature decreases from about $232$ to $73$\,K from the inner to the outer radius. Our best-fit models give a total dust mass of $(1.3\pm0.15)\times10^{-2}$\,M${_\oplus}$ within a radius of $1785$\,au; the dust mass estimate has a conservative uncertainty of about $\pm25$\%.

\item The material in the dust cloud may have come from a pre-existing Oort cloud analog, or it may lie in a disk-like structure produced as a result of binary interaction.

\item The radial extent of the dust cloud is almost identical to the orbital radius of the CS-A2V binary stellar system; the cloud appears flattened on the side facing the A2V star. These features suggest a physical interaction between the disk and the A2V companion, due to a combination of radiation pressure (due to the A2V star's radiation), together with partial destruction of the dust grains by sputtering (due to a tenuous wind from the A2V star).

\item A plausible evolutionary scenario that explains the spectacular mass-ejection that has resulted in the current, multipolar planetary nebula, is one in which three stars -- the CS, a close low-mass companion (with, say, $a\lesssim5$\,au), and a much more distant A2V star (with, say, $a\sim$400\,au) -- formed a stable hierarchical triple system on the main-sequence, but which then became dynamically active much later due to the Eccentric Kozai-Lidov mechanism causing a strong binary interaction between the inner pair and leading to a loss of most of the primary's envelope. The resulting severe reduction in the CS mass caused the A2V star's orbit to expand, resulting in its current separation from the CS.

\item We set an upper limit of $\sim$0.1\,\ms~on the mass of any main-sequence star, i.e., spectral-type M6V (L$\sim7\times10^{-4}$\,\ls, $T_{eff}\sim2500$\,K) or later, that may be located close enough to the CS to be unresolved and faint enough to be undetectable. Such a star could be the close binary companion in the above evolutionary scenario, provided the binary interaction did not lead to its merger with the CS.

\end{enumerate}

\section{acknowledgements}
We thank an (anonymous) referee for his/her timely and thorough review which has helped us improve our paper.

The Early Release Observations and associated materials were developed, executed, and compiled by the ERO production team:  Hannah Braun, Claire Blome, Matthew Brown, Margaret Carruthers, Dan Coe, Joseph DePasquale, Nestor Espinoza, Macarena Garcia Marin, Karl Gordon, Alaina Henry, Leah Hustak, Andi James, Ann Jenkins, Anton Koekemoer, Stephanie LaMassa, David Law, Alexandra Lockwood, Amaya Moro-Martin, Susan Mullally, Alyssa Pagan, Dani Player, Klaus Pontoppidan, Charles Proffitt, Christine Pulliam, Leah Ramsay, Swara Ravindranath, Neill Reid, Massimo Robberto, Elena Sabbi, Leonardo Ubeda. The EROs were also made possible by the foundational efforts and support from the JWST instruments, STScI planning and scheduling, and Data Management teams.

Most of the data presented in this paper were obtained from the Mikulski Archive for Space Telescopes (MAST) at the Space Telescope Science Institute. The specific observations analyzed can be accessed via \dataset[http://dx.doi.org/10.17909/6wd1-0170]{http://dx.doi.org/10.17909/6wd1-0170}. This work is also based in part on observations made with the Spitzer Space Telescope, which is operated by the Jet Propulsion Laboratory, California Institute of Technology under a contract with NASA, that can be accessed via  \dataset[https://doi.org/10.26131/irsa3]{https://doi.org/10.26131/irsa3}.

RS’s contribution to the research described here was carried out at the Jet Propulsion Laboratory, California Institute of Technology, under a contract with NASA. VB acknowledges support from the EVENTs/NEBULAE WEB research program, Spanish AEI grant PID2019-105203GB-C21. CSC's work is part of the I+D+i project PID2019-105203GB-C22 funded by the Spanish MCIN/AEI/10.13039/501100011033. The TMAW tool (http://astro.uni-tuebingen.de/~TMAW) used for this paper was constructed as part of the activities of the German Astrophysical Virtual Observatory.

\clearpage
\vskip 0.3in
\setlength{\bibsep}{3pt plus 0.3ex}

\begin{table*}
\caption{Photometry of the Central Star of NGC\,3132}
\label{tbl-flux-cs}
\begin{tabular}{llccccc}
\hline     
Filter & Wavelength & Flux   & Error\tablenotemark{a} & Apert.     & Apert. & Phot  \\
       & (\micron)  & (mJy)  & (\%)                   & Rad.($''$) & Corr.  & Ref.\tablenotemark{b} \\
\hline
U      & 0.36  & 2.15   &  15 &... & ...  & 14.8\tablenotemark{c}   (1)\\ 
F438W  & 0.438 & 1.80   &  15 & ... & ... & 15.76\tablenotemark{c}  (3)\\
F438W  & 0.438 & 1.62   &  15 & ... & ... & 15.876\tablenotemark{d} (2)\\ 
F555W  & 0.555 & 1.45   &  15 & ... & ... & 16.0\tablenotemark{c}   (3)\\   
F555W  & 0.555 & 1.19   &  15 & ... & ... & 16.212\tablenotemark{d} (2)\\ 
G-band & 0.639 & 1.18  &  15 & ... & ... & (4)\\
F814W  & 0.814 & 0.56  &  15 & ... & ... & 17.03\tablenotemark{d}   (3)\\
F090W  & 0.90  & 0.38   &  15 & $0\farcs10$ & 0.76 & (5)\\
F187N  & 1.87  & 0.11   &  15 & $0\farcs10$ & 0.75 & (5)\\
F212N  & 2.12  & 0.106  &  15 & $0\farcs10$ & 0.74 & (5)\\
F356W  & 3.56  & 0.037  &  15 & $0\farcs15$ & 0.71 & (5)\\
F405N  & 4.05  & 0.020  &  15 & $0\farcs15$ & 0.73 & (5)\\
F470N  & 4.70  & 0.018  &  15 & $0\farcs15$ & 0.72 & (5)\\
F770W  & 7.7   & 0.070  &  10 & $0\farcs35$ & 0.76 & (5)\\
F1130W & 11.3  & 1.16   &  10 & $0\farcs75$ & 0.97 & (5)\\
F1280W & 12.8  & 1.2    &  10 & $0\farcs75$ & 0.93 & (5)\\
F1800W & 18.0  & 9.9    &  10 & $1\farcs1\times1\farcs3$    & ... & (5)\\
F1800W & 18.0  & 10.3\tablenotemark{e}   &  10 & $1\farcs6$ & ... & (5)\\
F1800W & 18.0  & 11.1\tablenotemark{e}   &  10 & $1\farcs8$ & ... & (5)\\
F1800W & 18.0  & 12.1\tablenotemark{e}   &  10 & $2\farcs0$ & ... & (5)\\

\hline
\end{tabular}\\

\tablenotetext{a}{Percentage Error in Flux in previous column}
\tablenotetext{b}{References for photometry: (1) Kohoutek \& Laustsen 1977, (2) photometry on UVIS/WFC3 images (HST proposal 11699) from Monreal-Ibero \& Walsh (2020),  (3) Hubble Source Catalog V.3 (Whitmore et al. 2016), (4) $Gaia$ DR3, (5) this work; when pohotometry reference provides magnitudes, these are listed here}
\tablenotetext{c}{Vega Magnitude}
\tablenotetext{d}{AB Magnitude}
\tablenotetext{e}{Flux derived from integration of radial intensity to outer radius in Col.\,(3)}
\end{table*}

\begin{table*}
\caption{Photometry of Field Stars}
\label{tbl-flux-fs}
\begin{tabular}{lccccccccc}
\hline     
Star & Wavelength\tablenotemark{a} & Flux\tablenotemark{b}  & Error\tablenotemark{c} & Wavelength\tablenotemark{d} & Flux\tablenotemark{e}   &  Error\tablenotemark{c} & Instr.\tablenotemark{f} & Publ.Phot.  \\
      & (\micron)                   & (mJy)                 & (\%) & (\micron)                   & (mJy)                   &  (\%)  &                         &Ref.\tablenotemark{g} \\
\hline
fs3 &  3.56  & 0.38  & 5  & 3.56 & 0.359  & 0.4 &IRAC\,1 & 1 \\ 
fs3 &  7.7   & 0.085 & 10 & 7.91 & 0.075  & 7  &IRAC\,4 & 1 \\  
%
%
fs4 &  7.7   & 0.22  & 5 & 7.91 & 0.27   & 2.5 &IRAC\,4 & 1 \\ 
fs5 &  7.7   & 3.9   & 5 & 7.91 & 4.6    & 0.2 &IRAC\,4 & 1 \\ 
\hline
\end{tabular}\\
\tablenotetext{a}{Wavelength of JWST filter-passband}
\tablenotetext{b}{Measured photometry from this study}
\tablenotetext{c}{Percentage Error in Flux in previous column}
\tablenotetext{d}{Wavelength of filter-passband for published IRAC photometry (\cite{Fazio04})}
\tablenotetext{e}{Published photometry}
\tablenotetext{f}{Instrument/ Detector for published photometry}
\tablenotetext{g}{Reference for published photometry: (1) The Spitzer (SEIP) source list (SSTSL2) \citep{2021yCat.2368....0S}}
\end{table*}

\begin{table}
\caption{}
\label{tbl-psf}
\begin{tabular}{c c c c c}
\hline\hline      

Instr. & Filter & Box\tablenotemark{a} & \%\,Error\tablenotemark{b} & G.\,smooth\tablenotemark{c} \\
       &        &  (px/$''$)           &  in PSF\,scale-factor     &  ($''$)   \\
\hline                   
NIRCam &   F090W    &50/1.54   &  0.16$\times 10^{-2}$ &0.05    \\ 
NIRCam &   F187N    &50/1.54   &  0.20$\times 10^{-2}$ &0.03    \\
NIRCam &   F212N    &200/6.17  &  0.13$\times 10^{-2}$ &0.02    \\
NIRCam &   F356W    &50/3.15   &  0.47$\times 10^{-2}$ &0.05    \\
NIRCam &   F405N    &250/15.75 &  0.15$\times 10^{-1}$ &0.05    \\
NIRCam &   F470N    &50/3.15   &  0.17$\times 10^{-1}$ &0.045    \\
MIRI   &   F770W    &200/22.18 &  0.26$\times 10^{-1}$ &0.045    \\
MIRI   &   F1130W   &200/22.18 &  0.97$\times 10^{-1}$ &0.07    \\
MIRI   &   F1280W   &169/18.74 &  0.15 &0.1    \\
MIRI   &   F1800W   &135/14.97 &  0.49 &0.1    \\
\hline
\end{tabular}
\tablenotetext{a}{Box size used for estimating a median background}
\tablenotetext{b}{Percentage error in the scale-factor applied to the PSF for subtraction from A2V star image}
\tablenotetext{c}{FWHM of gaussian smoothing function applied to PSF before PSF-subtraction}
\end{table}
\begin{table}
\caption{Position Angles and Separations of Field Star Pairs}
\label{tbl-pa-sep}
\begin{tabular}{c c c c c c}
\hline\hline      

StarPair & $PA(HST)$\tablenotemark{a} & $PA($Gaia$,Epoch1)$\tablenotemark{b} & Ang.Dist.(HST)\tablenotemark{c} & Err(HST)\tablenotemark{d} &Frac.Diff.\tablenotemark{e}\\
         & \arcdeg & \arcdeg         & ${''}$ & ${''}$ & $10^{-5}$\\
\hline                   
fs6, fs1 &  22.016  & 21.850  & 98.1236  & 0.00055  &-2.58  \\ 
fs6, fs8 &  21.219  & 21.071  & 162.7896 & 0.0012  &-4.25  \\
fs1, fs8 &  20.031  & 19.912  & 64.6894  & 0.0011  &-6.36  \\
fs4, fs8 &  13.816  & 13.674  & 141.6898 & 0.0017  &-3.55  \\
fs4, fs1 &  8.647   &  8.486  & 77.6963  & 0.0013  &-2.85  \\
\hline
\end{tabular}
\tablenotetext{a}{PA of separation vector between star pair in Epoch 1, using HST image. PA is measured from N towards E, with the second star in the pair as the origin}
\tablenotetext{b}{PA of separation vector between star pair in Epoch 1, estimated from $Gaia$ DR3 data}
\tablenotetext{c}{Angular distance between star pair in Epoch 1, using HST image}
\tablenotetext{d}{Error in angular distance between star pair in Epoch 1, using HST image}
\tablenotetext{e}{Angular distance between star pair in Epoch 1 (using HST image) minus the angular distance between star pair in Epoch 1 (estimated from $Gaia$ DR3 data), divided by the average angular distance between star pair}
\end{table}

\begin{table}
\caption{Models of the Dust Emission towards the Central Star of NGC\,3132}
\label{mod-tbl}
\begin{tabular}{lllllll}
\hline
T$_{\rm d}$(in)\tablenotemark{a} & $R_{in}$\tablenotemark{b} &  $R_{out}$\tablenotemark{c} &  n\tablenotemark{d}   & $\tau _V$\tablenotemark{e} &  $F_{\rm bol}$\tablenotemark{f}  & M$_{\rm d}$\tablenotemark{g} \\
(K)                              & (arcsec, au)              & (arcsec, au)                &      &                         & (\fbolunit)     & (M${_\oplus}$) \\
\hline
\multicolumn{7}{l}{1-shell model}\\
\hline
232                            & $0\farcs095$, 71   &  $\gtrsim2\farcs37$, 1785           & -0.05 & $9.9\times10^{-4}$  & $6.5\times10^{-9}$ &  $1.15\times10^{-2}$ \\
\hline
\multicolumn{7}{l}{2-shell model} \\ 

\hline
232                           & $0\farcs095$, 71    &  $0\farcs47$, 357                   &  0.2  & $1.4\times10^{-4}$ & $6.5\times10^{-9}$ &  $7.4\times10^{-5}$ \\
125                           & $0\farcs47$, 357    &  $\gtrsim2\farcs37$, 1785           & -0.4  & $9.6\times10^{-4}$ & ...                &  $1.45\times10^{-2}$ \\
\hline
\end{tabular}
\tablenotetext{a}{The (input) dust temperature at shell inner radius}
\tablenotetext{b}{The (inferred) inner radius of the dust shell}
\tablenotetext{c}{The (inferred) outer radius of the dust shell}
\tablenotetext{d}{The (input) exponent of the density power law ($\rho_d(r)\propto r^{-n}$) in the dust shell}
\tablenotetext{e}{The (input) dust shell's optical depth at $0.55\micron$}
\tablenotetext{f}{Bolometric Flux}
\tablenotetext{g}{The (inferred) circumstellar dust mass}
\end{table}

%

\clearpage
\appendix
We show here the large-scale structure of the PN, NGC\,3132, imaged through the F1800W (18\,\micron) filter (Fig.\,\ref{pn-n3132-f1800w}). The image does not show the presence of any extended nebular feature (background or foreground) close to the compact dust clump around the central star (CS).

\renewcommand{\thefigure}{A.1\Alph{figure}}
\setcounter{figure}{0}
\begin{figure}[ht!]
\includegraphics[width=0.75\textwidth]{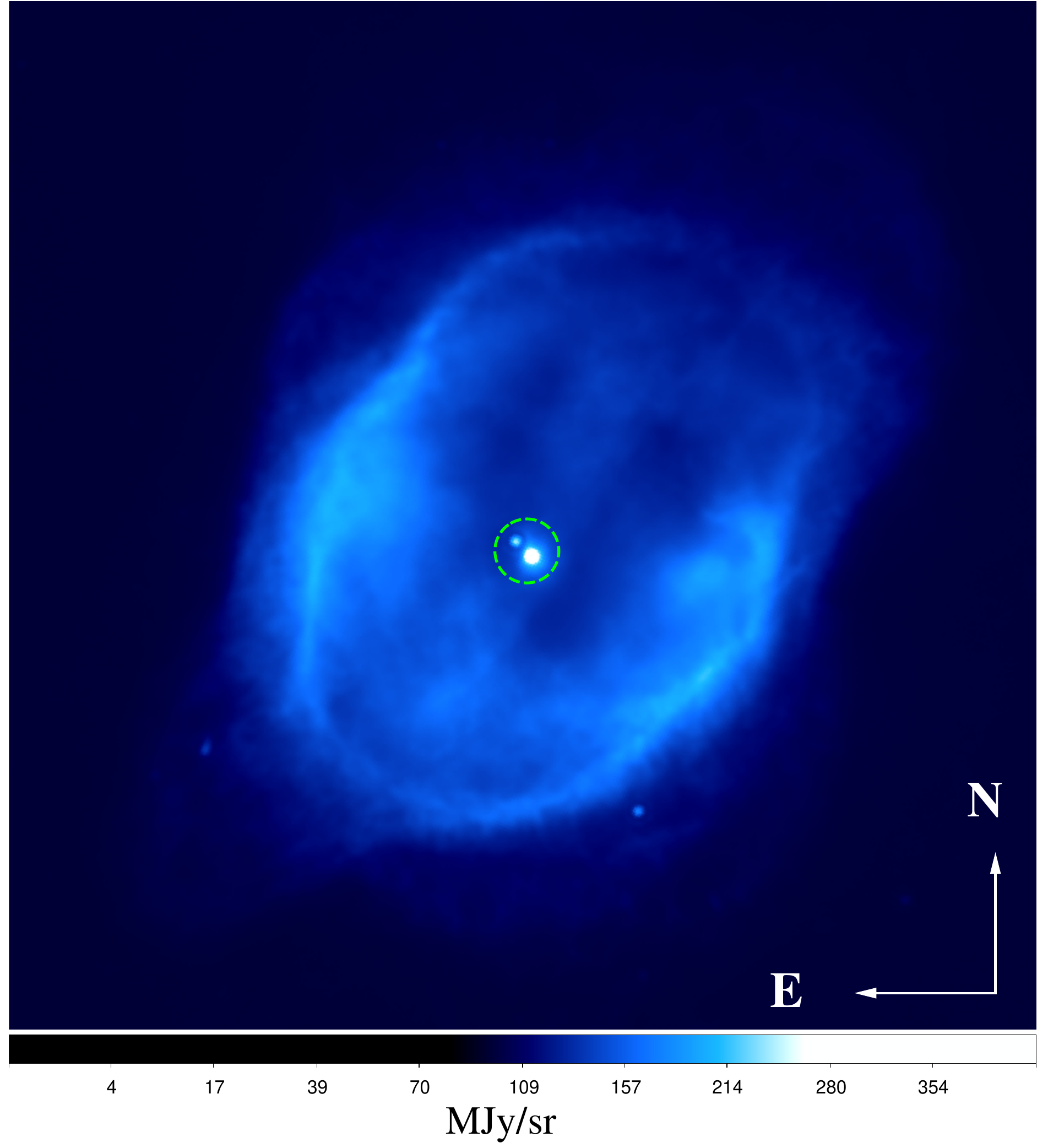}
\caption{The PN NGC\,3132 imaged with MIRI, through the F1800W (18\,\micron) filter, displayed using a square-root stretch. The green dashed circle (diameter $5{''}$) encloses the dust cloud around the CS, and the A2V star to its North-East. The panel size is $80{''}\times80{''}$.}
\label{pn-n3132-f1800w}
\end{figure}

\end{document}